\documentclass[conference]{IEEEtran}
\IEEEoverridecommandlockouts

\usepackage[noadjust]{cite}
\usepackage{cite}
\usepackage{amsmath,amssymb,amsfonts,dsfont}
\usepackage{algorithm}
\usepackage{algpseudocode}
\usepackage{graphicx}
\usepackage{textcomp}
\usepackage{xcolor}
\usepackage[Symbolsmallscale]{upgreek}
\usepackage{enumerate}
\usepackage{comment}
\usepackage[caption=false]{subfig}
\newcommand{\qedsymbol}{\hfill\square}
\usepackage{etoolbox}
\makeatletter
\patchcmd{\@makecaption}
  {\scshape}
  {}
  {}
  {}
\makeatletter
\patchcmd{\@makecaption}
  {\\}
  {.\ }
  {}
  {}
\makeatother

\newcommand{\R}{\mathbb{R}}
\newcommand{\N}{\mathbb{N}}

\newcommand{\Char}{\mathds{1}}
\newcommand{\ones}{\mathbf{1}}

\newcommand{\bfdelta}{\boldsymbol \delta}

\newtheorem{theorem}{Theorem}[section]

\newtheorem{proposition}[theorem]{Proposition}
\newtheorem{lemma}[theorem]{Lemma}

\newtheorem{definition}[theorem]{Definition}
\newtheorem{remark}[theorem]{Remark}

\newcommand{\mM}{\mathbf{M}}

\newcommand{\mD}{\mathbf{D}}

\newcommand{\mJ}{\mathbf{J}}
\newcommand{\mQ}{\mathbf{Q}}
\newcommand{\mP}{\mathbf{P}}

\newcommand{\mS}{\mathbf{S}}
\newcommand{\mA}{\mathbf{A}}
\newcommand{\mB}{\mathbf{B}}
\newcommand{\mC}{\mathbf{C}}

\newcommand{\eye}{\mathbf{I}}

\newcommand{\0}{\mathbf{0}}

\newcommand{\vx}{\mathbf{x}}
\newcommand{\vy}{\mathbf{y}}

\newcommand{\vr}{\mathbf{r}}
\newcommand{\vs}{\mathbf{s}}
\newcommand{\vecv}{\mathbf{v}}
\newcommand{\vw}{\mathbf{w}}
\newcommand{\vu}{\mathbf{u}}

\newcommand{\cG}{\mathcal{G}}
\newcommand{\cN}{\mathcal{N}}

\newcommand{\cE}{\mathcal{E}}

\begin{document}

\title{Competing Epidemics on Graphs - Global Convergence and Coexistence
\thanks{{\rule{5cm}{2pt}}

Vishwaraj Doshi is with the Operations Research Graduate Program, Shailaja Mallick is with the Department of Computer Science, and Do Young Eun is with the Department of Electrical and Computer Engineering, North Carolina State University, Raleigh, NC. Email: \{vdoshi, smallic, dyeun\}@ncsu.edu. This work was supported in part by National Science Foundation under Grant Nos. CNS-2007423 and CNS-1824518.}
}
\author{\IEEEauthorblockN{Vishwaraj Doshi}
\and
\IEEEauthorblockN{Shailaja Mallick}
\and
\IEEEauthorblockN{Do Young Eun}
}

\maketitle
\begin{abstract}
The dynamics of the spread of contagions such as viruses, infectious diseases or even rumors/opinions over contact networks (graphs) have effectively been captured by the well known \textit{Susceptible-Infected-Susceptible} ($SIS$) epidemic model in recent years. When it comes to competition between two such contagions spreading on overlaid graphs, their propagation is captured by so-called \textit{bi-virus} epidemic models. Analysis of such dynamical systems involve the identification of equilibrium points and its convergence properties, which determine whether either of the viruses dies out, or both survive together. We demonstrate how the existing works are unsuccessful in characterizing a large subset of the model parameter space, including all parameters for which the competitiveness of the bi-virus system is significant enough to attain coexistence of the epidemics. In this paper, we fill in this void and obtain convergence results for the entirety of the model parameter space; giving precise conditions (necessary and sufficient) under which the system \textit{globally converges} to a \textit{trichotomy} of possible outcomes: a virus-free state, a single-virus state, and to a coexistence state -- the first such result.
\end{abstract} 
\section{Introduction and overview}\label{introduction}
Graph based epidemic models are widely employed to analyze the spread of real-world phenomena such as communicable diseases~\cite{Yorke1976, hethcote2000mathematics}, computer viruses, malware~\cite{garetto2003modeling, yang2013epidemics, hosseini2016model}, product adoption~\cite{apt2011diffusion, prakash2012winner, ruf2017dynamics}, opinions and rumors~\cite{trpevski2010model, zhao2013sir, lin2018opinion}. The propagation of such phenomenon (which we cumulatively refer to as \textit{epidemics} or \textit{viruses}) usually takes place via processes such as human contact, word-of-mouth, exchange of emails or even in social media platforms. Graph based techniques, with edge based mechanisms to model information spread, have therefore proven to be effective in capturing such epidemic dynamics, and been a research focus over the past few decades \cite{shakkottaiINFOCOM2014, ganesh2005effect, draief_massoulie_2009}.

Epidemic models fall into various categories, with the main differentiating factor being whether the infected nodes remain infected for eternity or recover over time. In this paper, we focus on epidemics of the \textit{Susceptible-Infected-Susceptible} (SIS) type, where an infected node gradually recovers and becomes susceptible to disease once again. This model was originally introduced in \cite{Yorke1976} to capture the spread of Gonorrhea due to contact between individuals in a population, and was further developed in \cite{omic2009epidemic, van2009virus, gray2011stochastic, li2012susceptible, wang2012global, guo2013epidemic}. The central result for SIS epidemics is a \textit{dichotomy} arising from the relation between model parameter ($\tau\!>\!0$) representing the effective infection rate or strength of the virus,\footnote{$\tau = \beta/\delta$, where $\beta>0$ stands for the infection rate of the virus and $\delta>0$ the recovery rate from the virus. Section \ref{epidemic models} provides a detailed explanation.} and a threshold value ($\tau^*\!>\!0$). When $\tau \!\leq\! \tau^*$, the virus spread is not strong enough and the system converges to a `virus-free' state. When $\tau \!>\! \tau^*$, it converges to a state where the virus infects a non-zero portion of the population.


In recent years, there has been a strong interest in modelling the behaviour of two competing products in a market, the clash of conflicting rumors, opposing opinions in the real-world, or competing memes on social platforms; in general the dynamics of two competing `epidemics' via graph based models \cite{apt2011diffusion, wang2012dynamics}. The key questions posed in such literature are: Can both competing epidemics coexist over the network? If not, which one prevails? Or do both die out? This \textit{trichotomy} of possible results is what recent literature has been trying to characterize, with many works especially focusing on the so-called \textit{bi-virus} or \textit{bi-SIS} model, which captures the spread of two competing SIS epidemics over a network \cite{prakash2012winner, Santos2015, yang2017bi, liu2019analysis}.

When the propagation of the two epidemics occurs over the same network \cite{prakash2012winner, wang2012dynamics}, it has been established that coexistence of two viruses is impossible except in the rare cases where their effective strengths ($\tau_1,\tau_2\!>\!0$ for viruses 1, 2, respectively) are equal \cite{liu2019analysis, prakash2012winner, yang2017bi, Santos2015, sahneh2014competitive}; the virus with the larger effective strength otherwise wiping out the other, a phenomenon sometimes referred to as \textit{winner takes all} \cite{prakash2012winner}. The situation is much more complicated when the two viruses spread over two distinct networks overlaid on the same set of nodes. This modelling approach is more representative of the real-world, where competing rumors/products/memes may not use the same platforms to propagate, though they target the same individuals. Recent works \cite{sahneh2014competitive, liu2019analysis, Santos2015, yang2017bi} therefore consider this more general setting, but unfortunately, a complete characterization of the trichotomy of outcomes has still proven to be elusive, and remains open as of now.

\subsubsection*{\textbf{Status Quo}} Of all recent works concerning the spread of SIS type bi-virus epidemics on overlaid networks, \cite{yang2017bi} and \cite{Santos2015} provide conditions under which the system globally converges to the state where one virus survives while the other dies out. Epidemics are typically modelled as systems of ordinary differential equations (ODEs), and \cite{yang2017bi} approaches the problem of showing global convergence by employing the classic technique via Lyapunov functions. However, finding appropriate Lyapunov functions is a highly non-trivial task, and as mentioned in \cite{Santos2015}, is even more difficult due to the coupled nature of bi-virus ODE systems. This can be seen in the condition they derive in \cite{yang2017bi} for the case where, say, Virus 1 dies out and Virus 2 survives, since it contains the term $\tau_1 \!\leq\! \tau_1^*$ in it (where $\tau_1^*$ is the threshold corresponding to the single-virus case), meaning that Virus 1 would not have survived even if it was the only epidemic present on the network. More importantly, \cite{yang2017bi} is unable to characterize convergence properties for $\tau_1\!>\!\tau_1^*$ and $\tau_2\!>\!\tau_2^*$.

The authors in \cite{Santos2015} take a different approach and tackle this problem by applying their `qualitative analysis' technique, which uses results from other dynamical systems that bound the solutions of the bi-virus ODE; and provide conditions under which the system globally converges to single-virus equilibria. As we show later in Section \ref{convergence and coexistence}, however, their conditions not only characterize just a \textit{subset} of the actual space of parameters that lead to global convergence to the single-virus equilibria (which they themselves pointed out), but the size of this subset is highly sensitive to the graph topology, often much smaller than what it should be in general. In other words, a complete characterization of the \emph{entire} space of model parameters, on which the system globally converges to one of the trichotomic states, has still been recognized as an open problem in the bi-virus literature \cite{yang2017bi, Santos2015, liu2019analysis}.

\subsubsection*{\textbf{Our Contributions}} 
In this paper, we provide the complete characterization of the trichotomy of the outcomes with necessary and sufficient conditions on the model parameters under which the system globally converges to one of the three possible points: (i) a `virus-free' state, (ii) a single-virus equilibrium (winner takes all), or (iii) an equilibrium where both viruses coexist over the network. While the result for the virus-free state is not new and stems directly from the classical result for single-virus SIS models from \cite{Yorke1976} as demonstrated in \cite{yang2017bi, Santos2015 ,liu2019analysis}, the convergence results for the other two types of outcomes are not straightforward, rendering the typical Lyapunov based approach largely inapplicable. 

In proving these results, we first show, using a specially constructed cone based ordering, that the bi-virus epidemic model possesses some inherent monotonicity properties. We then use novel techniques from the theory of \textit{monotone dynamical systems} (MDS) to prove our main results. We note that our characterization of the parameter space coincides with that in \cite{sahneh2014competitive}, wherein they employed only \textit{local} stability results via bifurcation analysis -- concerning only solution trajectories that originate from a small neighborhood of those fixed points. In contrast, in this paper, we show the \emph{global} stability of the system for any possible combination of the model parameters, starting from any initial point. To the best of our knowledge, our work is the first one that establishes the \emph{complete} characterization of the bi-virus system with global convergence, thereby solving the aforementioned open problems in the bi-virus literature.

\subsubsection*{\textbf{Structure of the paper}}
In Section \ref{epidemic models}, we first introduce the basic notation used throughout the paper, along with the classical (single-virus) SIS model and the bi-virus model. In Section \ref{monotonicity of epidemic models}, we provide a primer to the MDS theory, which we then utilize to show that the bi-virus epidemic model is a monotone dynamical system. We include the main convergence results in Section \ref{convergence and coexistence}, along with brief comparisons with existing literature. In Section \ref{numerical results}, we provide numerical results which confirm our theoretical findings. We then conclude in Section \ref{conclusion}. For better readability of the paper, all technical proofs of main results are deferred to Appendix \ref{proof of the results}. The appendices also include some selected definitions and results from matrix theory (Appendix \ref{matrix theory results}), ODE theory (Appendix \ref{ODE results}), and from MDS theory (Appendix \ref{MDS}), which we use as part of our proofs of the Theorems in Section \ref{convergence and coexistence}.
\section{Preliminaries} \label{epidemic models}

\subsection{Basic Notations}
We standardize the notations of vectors and matrices by using lower case, bold-faced letters to denote vectors ($\vecv \!\in\! \R^{N}$), and upper case, bold-faced letters to denote matrices ($\mM \!\in\! \R^{N \times N}$). We denote by $\lambda(\mM)$ the largest \textit{real part}\footnote{We use the $\lambda$ notation instead of something like $\lambda_{Re}$, since it will mostly be used in cases where the largest eigenvalue is real, for which $\lambda$ itself is the largest real eigenvalue. For example, $\lambda(\mA)$ becomes the spectral radius for any non-negative matrix $\mA$ \cite{meyer_textbook}.} of all eigenvalues of a square matrix $\mM$. We use $\text{diag}(\vecv)$ or $\mD_{\vecv}$ to denote the $N \!\!\times\!\! N$ diagonal matrix with entries of vector $\vecv \in \R^N$ on the diagonal. Also, we denote $\ones \!\triangleq\! [1,\!\cdots\!,1]^T$ and $\0 \!\triangleq\! [0,\!\cdots\!,0]^T$, the $N$-dimensional vector of all ones and zeros, respectively. For vectors, we write $\vx\!\leq\!\vy$ to indicate that $x_i \!\leq\! y_i$ for all $i$; $\vx \!<\! \vy$ if $\vx\!\leq\!\vy$ and $\vx\!\neq\!\vy$; $\vx \!\ll\! \vy$ when all entries satisfy $x_i \!<\! y_i$. We use $\cG(\cN,\cE)$ to represent a general, undirected, connected graph with $\cN \triangleq \{1,2,\cdots, N\}$ being the set of nodes and $\cE$ being the set of edges. When we refer to a matrix $\mA \!=\! [a_{ij}]$ as the adjacency matrix of some graph $\cG(\cN,\cE)$, it satisfies $a_{ij} \triangleq \Char_{\lbrace(i,j) \in \cE\rbrace}$ for any $i,j \in \cN$; we use $d_{min}(\mA)$ and $d_{max}(\mA)$ to denote the minimum and maximum degrees of the nodes of the corresponding graph. Since we only consider connected graphs, all the adjacency matrices in this paper are automatically considered to be irreducible (see Definition \ref{irreducible matrix} in Appendix \ref{matrix theory results}).
\subsection{The $SIS$ Model}\label{sis epidemic model}
Consider the graph $\cG(\cN,\cE)$, and assume that at any given time $t \geq 0$, each node $i \in \cN$ of the graph is either in an \emph{infected (I)}, or in a \emph{susceptible (S)} state. An infected node can infect each of its susceptible neighbors with rate $\beta>0$.\footnote{We say an event occurs with some \emph{rate} $\alpha>0$ if it occurs after a random amount of time, exponentially distributed with parameter $\alpha>0$.} It can also, with rate $\delta>0$, be cured from its infection and revert to being susceptible again. We write $\vx(t) = [x_i(t)] \in \R^N$, where $x_i(t)$ represents the probability that node $i\in\cN$ is infected at any given time $t\geq0$. Then, the dynamics of the $SIS$ model can be captured via the system of ODEs given by
\begin{equation}\label{ode-sis-i}
    \frac{d x_i(t)}{dt} = f_i(\vx(t)) \triangleq \beta (1-x_i(t)) \sum_{j \in \cN} a_{ij} x_j(t) - \delta x_i(t)
\end{equation}
\noindent for all $i \in \cN$ and $t \geq 0$. In a matrix-vector form, this can be written as
\begin{equation}\label{ode-sis}
  \frac{d\vx}{dt} = F(\vx) \triangleq \beta \text{diag} (\ones - \vx)\mA\vx - \delta \vx
\end{equation}
\noindent where we suppress the $(t)$ notation for brevity. The system \eqref{ode-sis} is positively invariant in the set $[0,1]^N$, and has $\0$ as one of its fixed points (the virus-free equilibrium). The following results are well known from \cite{Yorke1976}, and will be critical for the proofs of our main results in Section \ref{convergence and coexistence}.\linebreak

\begin{theorem}[Theorem 3.1 in \cite{Yorke1976}] \label{sis-conditions}
  Let $\tau \!\triangleq\! \beta/\delta$. Then,
  \begin{itemize}
    \item[(i)] either $\tau\leq 1/\lambda(\mA)$, and $\vx^* = \0$ is a globally asymptotically stable fixed point of \eqref{ode-sis};
    \item[(ii)] or $\tau > 1/\lambda(\mA)$, and there exists a unique, strictly positive fixed point $\vx^* \in (0,1)^N$ such that $\vx^*$ is globally asymptotically stable in $[0,1]^N\setminus \{\0\}$.$\qedsymbol$
  \end{itemize}
\end{theorem}
\subsection{The Bi-Virus Model}\label{bi-virus epidemic model}

Consider two graphs $\cG_1(\cN,\cE_1)$ and $\cG_2(\cN,\cE_2)$, on the same set of nodes $\cN$ but with different edge sets $\cE_1$ and $\cE_2$. At any given time $t\geq 0$, a node $i \in \cN$ is either \textit{infected by Virus 1}, \textit{infected by Virus 2}, or is \textit{susceptible}. A node infected by Virus 1 infects each of its susceptible neighbors with rate $\beta_1 > 0$, just like in the $SIS$ model, but does so only to nodes which are its neighbors with respect to the graph $\cG_1(\cN,\cE_1)$. Nodes infected by Virus 1 also recover with rate $\delta_1>0$, after which they enter the susceptible state. Similarly, nodes infected by Virus 2 infect their susceptible neighbors, this time with respect to the graph $\cG_2(\cN,\cE_2)$, with rate $\beta_2>0$, while recovering with rate $\delta_2>0$. This competing bi-virus model of epidemic spread, also referred to as the $SI_1I_2S$ model, can be represented by the following ODE system:
\begin{equation}\label{eq:biSIS-i}
\begin{split}
    \frac{d x_{i}}{dt} = g_i(\vx,\vy) &\triangleq \beta_1 \left(1- x_i - y_i\right) \sum_{j \in \cN} a_{ij} x_{j} - \delta_1 x_{i} \\
    \frac{d y_{i}}{dt} = h_i(\vx,\vy) &\triangleq \beta_2 \left(1- x_i - y_i\right) \sum_{j \in \cN} b_{ij} y_{j} - \delta_2 y_{i}
\end{split}
\end{equation}
\noindent for all $i \in \cN$ and $t\geq 0$. In matrix-vector form, \eqref{eq:biSIS-i} becomes:
\begin{equation}\label{eq:biSIS}
\begin{split}
    \frac{d \vx}{dt} = G(\vx,\vy) &\triangleq \beta_1 \text{diag}\left(\ones - \vx - \vy\right) \mA \vx - \delta_1 \vx \\
    \frac{d \vy}{dt} = H(\vx,\vy) &\triangleq \beta_2 \text{diag}\left(\ones - \vx - \vy\right) \mB \vy - \delta_2 \vy,
\end{split}
\end{equation}
\noindent where $\mA = [a_{ij}]$ and $\mB = [b_{ij}]$ are the adjacency matrices of graphs $\cG_1(\cN,\cE_1)$ and $\cG_2(\cN,\cE_2)$, respectively.

\section{Monotonicity of epidemic models}\label{monotonicity of epidemic models}

In this section, we provide a succinct introduction to monotone dynamical systems and some important definitions therein. We then show that the bi-virus model \eqref{eq:biSIS} is a monotone dynamical system (specifically a cooperative system) with respect to some well-defined cone-orderings. 

\subsection{Monotone Dynamical Systems - A Primer}
A well known result from real analysis is that monotone sequences in compact (closed and bounded) subsets of $\R^n$ converge in $\R^n$ \cite{Analysis-Yeh}. This simple, yet powerful result has been fully integrated with the theory of dynamical systems in a series of works \cite{Hirsch-I,Hirsch-II,Hirsch-III,Hirsch-IV,Hirsch-V,HLSmith'88,HLSmith'90,HLSmith'91,HLSmith'04, MDSbook}, which cumulatively form the theory of \textit{monotone dynamical systems} (MDS). The foundations of MDS were laid down in \cite{Hirsch-I,Hirsch-II,Hirsch-III,Hirsch-IV,Hirsch-V} which study ordinary differential equations, specifically \emph{cooperative} ODE systems. We here provide a brief, informal introduction to such ODE systems, with more details in Appendix \ref{MDS}.

A central tool in the theory of MDS is the notion of \textit{generalized cone-orderings}, which extends the concept of monotonicity in vector spaces. Given a convex cone $K \subset X$ for any vector space $X$, the \textit{cone-ordering} $\leq_K$ ($<_K$, $\ll_K$) generated by $K$ is an order relation that satisfies
$\text{(i)} ~\vx \!\leq_K\! \vy \!\iff\! (\vy\!-\!\vx) \in K$;
$\text{(ii)} ~\vx \!<_K\! \vy \!\iff\! \vx \!\leq_K\! \vy$ and $\vx \!\neq\! \vy$; and
$\text{(iii)} ~\vx \!\ll_K\! \vy \!\iff\! (\vy\!-\!\vx) \in \text{int}(K)$, for any $\vx,\vy \in X$. Note that, `$\ll_K$' implies `$<_K$' and is a stronger relation. Cone-orderings generated by the positive orthant $K \!=\! \R^n_+$ are simply denoted by $\leq$ ($<, \ll$), that is, without the `$K$' notation.

Let $\phi_t(\vx)$ denote the solution of a dynamical system at some time $t \!>\! 0$ starting from an initial point $\phi_0(\vx) \!=\! \vx \!\in\! \R^n$. Then given a cone-ordering $\leq_K$ ($<_K$, $\ll_K$), the dynamical system is said to be \emph{monotone} if for every $\vx,\vy \!\in\! \R^n$ such that $\vx \!\leq_K\! \vy$, we have $\phi_t(\vx) \!\leq_K\! \phi_t(\vy)$ for all $t \!>\! 0$. The system is called \emph{strongly monotone} if for all $\vx,\vy \!\in\! \R^n$ such that $\vx \!<_K\! \vy$, we have $\phi_t(\vx) \!\ll_K\! \phi_t(\vy)$ for all $t \!>\! 0$. The main result from MDS theory says that (almost) every solution trajectory of a \textit{strongly monotone} system always converges to some equilibrium point of the system \cite{Hirsch-II,HLSmith'91,HLSmith'04,HLSmith'17}. If the system has only one stable fixed point, then this in itself is enough to prove global convergence. Monotonicity properties of a dynamical system can therefore be leveraged as an alternative to constructing Lyapunov functions, which is often intractable.

Consider the following autonomous ODE system
\begin{equation}\label{ode sys}
  \dot \vx  = F(\vx),
\end{equation}
\noindent where $F(\vx) = [f_i(\vx)] \in \R^n$ is the vector field. If $\phi_t(\vx)$ is the solution of this ODE system, we say the system is \emph{co-operative} if it is monotone. There are ways to find out whether an ODE system is co-operative or not. In particular, one can answer this by observing the Jacobian of the vector field \cite{HS_Co-Op}. The so-called \textit{Kamke condition} \cite{MDSbook} says that \eqref{ode sys} is co-operative with respect to the cone-ordering generated by the positive orthant $K=\R^n_+$ if and only if 
\begin{equation}\label{Kamke for positive}
  \frac{\partial f_i}{\partial x_i} \geq 0, ~~~~~~~ \text{for all } i \neq j.
\end{equation}
\noindent While it is not straightforward to obtain such a clean condition for any general convex cone $K$, one can still deduce the co-operative property of the ODE with respect to any one of the other orthants of $\R^n$ by observing the signed entries of the Jacobian. We will show how this is done for the bi-virus system \eqref{eq:biSIS} later in Section \ref{monotonicity bi-virus}.

If the Jacobian of an ODE system is an irreducible matrix in a subset $D$ of the state space, we say that the ODE system is \textit{irreducible in $D$} (Definition \ref{irreducible ode} in Appendix \ref{MDS}). If the ODE system is co-operative in $D$ as well as irreducible in $D$, then it is strongly monotone in $D$ (Theorem \ref{SM ODE} in Appendix \ref{MDS}). To prove convergence properties, we should ideally be able to show that our system is strongly monotone in the entirety of the state space it is contained in, for which we can directly apply the main MDS convergence result. However, this is often not the case, and one needs additional results from MDS literature to prove convergence. These details are deferred to Appendix \ref{MDS}.

\subsection{Monotonicity of the Bi-Virus Model} \label{monotonicity bi-virus}

We now turn our attention to competing epidemics and establish monotonicity results about the bi-virus model.

\subsubsection*{\textbf{Southeast cone-ordering and the Kamke condition}}
Consider the cone-ordering generated by the convex cone $K=\{ \R^N_+ \times \R^N_- \} \subset \R^{2N}$. This cone is one of the orthants of $\R^{2N}$, and for $N=1$, it would correspond to the \textit{southeast} orthant of $\R^2$ $\left( K = \{ \R_+ \times \R_- \} \subset \R^2 \right)$. For any two points $(\vx,\vy)$, $(\bar \vx, \bar \vy) \in \R^{2N}$, it satisfies the following:
\begin{itemize}
\item[(i)]
$(\vx,\vy) \!\leq_K\! (\bar \vx,\bar \vy) \!\iff\! x_i \!\leq\! \bar x_i$ and $y_i \!\geq\! \bar y_i$ for all $i \!\in\! \cN$;
\item[(ii)]
$(\vx,\vy) \!\!<_K\!\! (\bar \vx,\bar \vy) \!\iff\! (\vx,\vy) \!\!\leq_K\!\! (\bar \vx,\bar \vy)$ and $(\vx,\vy) \!\!\neq\!\! (\bar \vx,\bar \vy)$;
\item[(iii)]
$(\vx,\vy) \!\ll_K\! (\bar \vx,\bar \vy) \!\iff\! x_i \!<\! \bar x_i$ and $y_i \!>\! \bar y_i$ for all $i \!\in\! \cN$.
\end{itemize}
This type of cone-ordering is often referred to as the \textit{southeast cone-ordering}, and the corresponding cone $K$ is the \textit{southeast orthant} of $\R^{2N}$. As shown in \cite{HS_Co-Op}, the Kamke condition for determining whether an ODE system is cooperative or not with respect to the positive orthant $\R^{2N}_+$ can be generalised for cone-orderings generated by any orthant of $\R^{2N}$, including the southeast orthant. Once again, this is done by observing the Jacobian of the respective ODE system. Consider the $2N$ dimensional system given by\vspace{-1mm}
\begin{equation*}\vspace{-1mm}
  \dot \vx = G(\vx,\vy) ~~\text{and}
  ~~\dot \vy = H(\vx,\vy),
\end{equation*}
\noindent where $G(\vx,\vy) = [g_i(\vx,\vy)]$ and $H(\vx,\vy) = [h_i(\vx,\vy)]$ are vector-valued functions in $\R^N$. The Kamke condition for this system with respect to the southeast cone-ordering \cite{HS_Co-Op} is
\begin{equation*}
    \frac{\partial g_i}{\partial x_j} \geq 0, ~ \frac{\partial h_i}{\partial y_j} \geq 0, ~\forall i\neq j, ~~~\text{and}~~~
\frac{\partial g_i}{\partial y_j} \leq 0, ~ \frac{\partial h_i}{\partial x_j} \leq 0, ~\forall i,j.
\end{equation*}
\noindent Roughly speaking, the Jacobian $\mJ_{GH} (\vx,\vy)$ of the system, evaluated at all points in the state space, should be in the following block matrix form (where the signs are not strict):
\begin{equation}\label{Jacobian form} 
\mJ_{GH}=
\begin{bmatrix}
*  &+  &+  &- &- &- \\
+  &*  &+  &- &- &- \\
+  &+  &*  &- &- &- \\
-  &-  &-  &* &+ &+ \\
-  &-  &-  &+ &* &+ \\
-  &-  &-  &+ &+ &*
\end{bmatrix}
\end{equation}
\noindent  Note that the state space of the ODE system \eqref{eq:biSIS} is given by $D \triangleq \left\{ (\vx,\vy) \in [0,1]^{2N} ~|~ \vx + \vy \leq \ones \right\}$.

\begin{proposition}\label{cooperative interor prop}
  The ODE system \eqref{eq:biSIS} (the bi-virus model) is cooperative in $D$ with respect to the southeast cone-ordering. It is also irreducible in $\text{Int}(D)$.$\qedsymbol$
\end{proposition}

\begin{IEEEproof}
  For all $(\vx,\vy) \in D$ and $i \neq j \in \cN$, we have
  $$\frac{\partial g_i}{\partial x_j} \!=\! \beta_1(1-x_i-y_i)a_{ij} \geq 0
  ~~\text{and}~~
  \frac{\partial h_i}{\partial y_j} \!=\! \beta_2(1-x_i-y_i)b_{ij} \geq 0$$
  \noindent since $a_{ij}\geq 0$, $b_{ij}\geq 0, (1-x_i-y_i)\geq 0$. Also for all $i\neq j$,
  $$\frac{\partial g_i}{\partial y_i} = -\beta_1[\mA\vx]_i \leq 0
  ~~\text{and}~~
  \frac{\partial h_i}{\partial x_i} = -\beta_2[\mB\vy]_i \leq 0,$$
  \noindent with ${\partial g_i}/{\partial y_j}={\partial h_i}/{\partial x_j}=0$. Thus, the Kamke conditions are satisfied and the system is cooperative in $D$.

  The Jacobian $\mJ_{GH} (\vx,\vy)$ of system \eqref{eq:biSIS} is written as
  \begin{equation}\label{Jacobian bi-virus}
    \begin{split}
      &\mJ_{GH}(\vx,\vy) = \\
      &\!\!\begin{bmatrix}
        \beta_1 \mS_{\vx\vy}\mA - \beta_1\mD_{\mA\vx} -\delta_1 \eye      &\!\!\!\!- \beta_1\mD_{\mA\vx}  \\
        - \beta_2\mD_{\mB\vy}    &\!\!\!\!\beta_2 \mS_{\vx\vy}\mB -  \beta_2\mD_{\mB\vy} -\delta_2 \eye
      \end{bmatrix},
    \end{split}
  \end{equation}
  \noindent where $\mS_{\vx,\vy} \triangleq \text{diag}(\ones - \vx - \vy)$, $\mD_{\mA\vx} \triangleq \text{diag}(\mA \vx)$ and $\mD_{\mB\vy} \triangleq \text{diag}(\mB\vy)$. Since $\mA$ and $\mB$ are irreducible, and the off-diagonal blocks are non-zero for $(\vx,\vy) \in \text{Int}(D)$, there does not exist a permutation matrix that would transform this into a block upper triangular matrix. Hence, by Definition \ref{irreducible ode}, the system is irreducible in $\text{Int}(D)$, and this completes the proof.
\end{IEEEproof}

From Proposition \ref{cooperative interor prop}, we deduce that the bi-virus system of ODEs \eqref{eq:biSIS} is co-operative in $D$, and thus strongly monotone in $\text{Int}(D)$ in view of Theorem \ref{SM ODE} in Appendix \ref{MDS}. 

\begin{remark} \label{remark sis coop}
When the Jacobian in \eqref{Jacobian bi-virus} is evaluated at any point of the type $(\vx,\0)$, the top left diagonal matrix reduces to $\beta_1 \text{diag}(\ones-\vx)\mA - \beta_1 \mD_{\mA\vx} - \delta_1 \eye$. This is exactly the Jacobian of a single-virus SIS system \eqref{ode-sis} with parameters $\beta_1$ and $\delta_1$ evaluated at any point $\vx\in[0,1]^N$. Since $\partial g_i/\partial x_j \geq 0$ even when $y_i\!=\!0$, the single-virus SIS system satisfies the Kamke conditions in \eqref{Kamke for positive}, and is thus a cooperative system, with respect to the cone-ordering generated by the positive orthant $\R^N_+$. Likewise, the irreducibility of the system in $(0,1)^N$ follows from that of $\mA$ as before. Hence, from Theorem \ref{SM ODE}, the single-virus SIS system is strongly monotone in $(0,1)^N$, with respect to $K\!=\!\R^N_+$. $\qedsymbol$
\end{remark} 
\section{Convergence and Coexistence properties of the Bi-Virus model}  \label{convergence and coexistence}

\begin{figure*}[!ht]
    \centering
    \vspace{-2mm}
    \hspace{-1mm}
    \subfloat[Limitations of the literature.]{\includegraphics[scale=0.6]{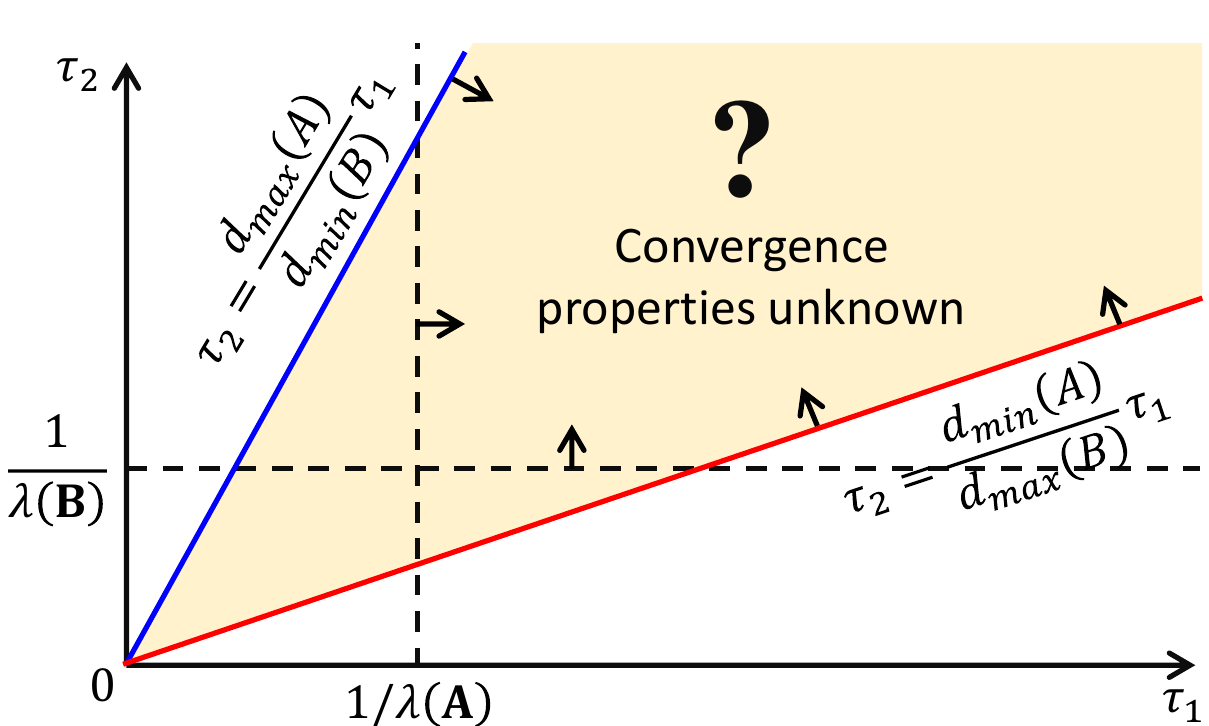}}
    \hfil
    \subfloat[Complete characterization of the convergence trichotomy.]{\includegraphics[scale=0.6]{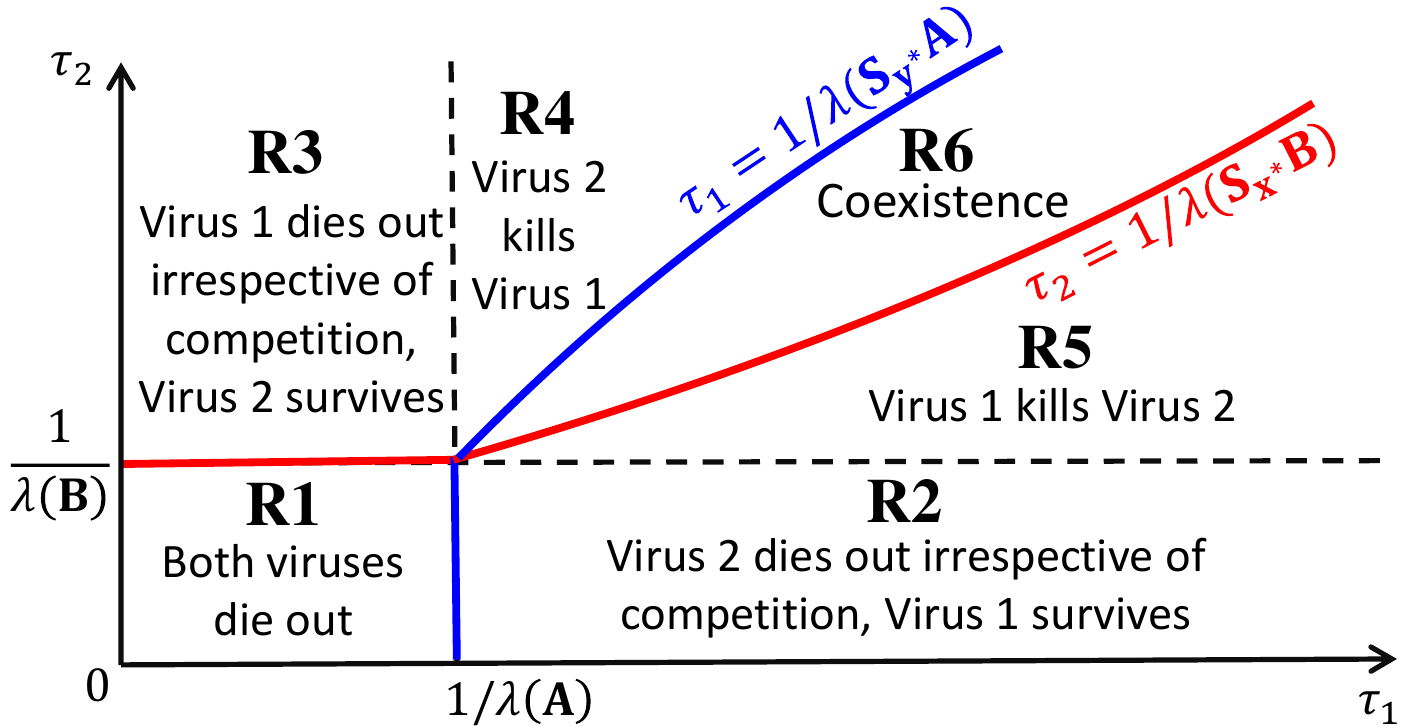}}
    \caption{Characterization of the parameter space}\vspace{-5mm}\label{all regions}
\end{figure*}

We are now ready to establish results on convergence properties of the bi-virus model and provide conditions for coexistence of two viruses in the bi-virus model as in \eqref{eq:biSIS}. We begin by defining some additional notation around fixed points of the bi-virus system. We defer all the proofs of our results in this section to Appendix \ref{proof of the results}.

Let $\vx^*$ and $\vy^*$ be the globally attractive fixed points of the single-virus SIS models that system \eqref{eq:biSIS} would reduce to when Virus 2 and 1, respectively, are not present over the network. These systems are given by
\begin{equation}\label{sis-x}
  \dot \vx = F^x(\vx) \triangleq \beta_1 \text{diag}(\ones - \vx)\mA \vx - \delta_1 \vx,
\end{equation}
\begin{equation}\label{sis-y}
  \dot \vy = F^y(\vy) \triangleq \beta_2 \text{diag}(\ones - \vy)\mB \vy - \delta_2 \vy;
\end{equation}
\noindent and by Theorem \ref{sis-conditions}, $\vx^*\!=\!\0$ ($\vy^*\!=\!\0$) if $\tau_1\lambda(\mA) \!\leq\! 1$ (if $\tau_2\lambda(\mB) \!\leq\! 1$), and $\vx^*\!\gg\!\0$ ($\vy^*\!\gg\!\0$) otherwise.

We first state the result when the virus-free equilibrium is globally attractive. We prove this by presenting simple arguments which require only Theorem \ref{sis-conditions} for SIS model along with the monotonicity properties derived in the previous section, eliminating the need of a Lyapunov based approach.
\begin{theorem}[Convergence to virus-free equilibria]\label{theorem virus free}
  If $\tau_1\lambda(\mA) \!\leq\! 1$ and $\tau_2\lambda(\mB) \!\leq\! 1$, trajectories of \eqref{eq:biSIS} starting from any point in $D$ converge to $(\0,\0)$.$\qedsymbol$
\end{theorem}

We next characterize the conditions when the system globally converges to equilibria when only one of the viruses survives over the network. Let $\mS_\vx \!\triangleq\! \text{diag}(\ones\!-\!\vx)$ and $\mS_\vy \!\triangleq\! \text{diag}(\ones\!-\!\vy)$ for any $\vx,\vy \in \R^N$. Also denote by $B_x \!\triangleq\! \left\{ (\vx,\vy) \in D ~|~ \vx\!>\!\0 \right\}$ the set of all points $(\vx,\vy) \!\in\! D$ for which $x_i\!>\!0$ for some $i\in\N$, and let $B_y \!\triangleq\! \left\{ (\vx,\vy) \in D ~|~ \vy\!>\!\0 \right\}$ be a similar set for the $y_i$ entries.\linebreak
\begin{theorem}[Convergence to single-virus equilibria]\label{theorem wta}
  When $\tau_1\lambda(\mS_{\vy^*} \mA ) \!>\! 1$ and $\tau_2\lambda(\mS_{\vx^*} \mB) \!\leq\! 1$, $(\vx^*,\0)$ is globally attractive in $B_x$;\footnote{We consider $B_x$ as the global domain of attraction instead of $D$ because $\vx=0$ for all points in the set $D\setminus B_x$. Starting from such points the system is no longer a bi-virus epidemic, but a single-virus SIS system for Virus 2.} that is, every trajectory of system \eqref{eq:biSIS} starting from points in $B_x$ converges to $(\vx^*,\0)$.

  Similarly, when $\tau_1\lambda(\mS_{\vy^*} \mA ) \!\leq\! 1$ and $\tau_2\lambda(\mS_{\vx^*} \mB) \!>\! 1$, $(\0,\vy^*)$ is globally attractive in $B_y$. $\qedsymbol$
\end{theorem}
\begin{figure}[!h]
    \centering
    \vspace{-6mm}
    \subfloat[For every point $p_k$, there is a point $(\vx_{rk}, \vy_{sk})$ starting from which, trajectories converge monotonically $(\leq_{K})$ to $(\vx^*,0)$.]{\includegraphics[scale=0.55]{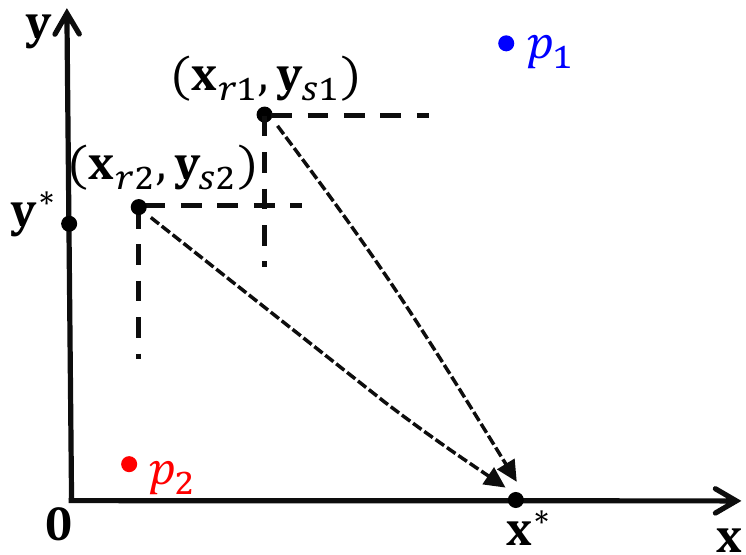}}
    \hfil
    \subfloat[Trajectories starting from $p_k$ eventually bounded by $(\vx_{rk}, \vy_{sk})$; monotonicity of the system gives convergence to $(\vx^*,0)$.]{\includegraphics[scale=0.55]{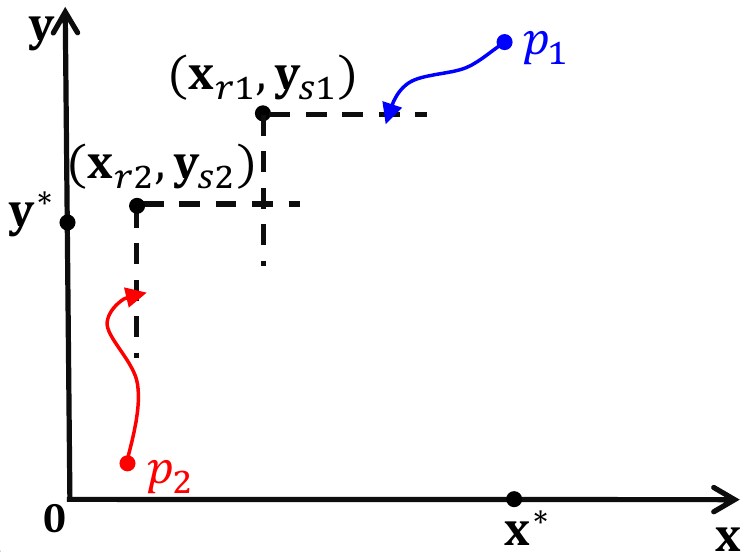}}
    \caption{Illustration of the convergence to $(\vx^*,0)$}\vspace{-2mm}\label{sketch}
\end{figure}

\begin{IEEEproof}[Sketch of the proof (convergence to $(\vx^*,\0)$)]
The idea behind the proof is illustrated in Figure \ref{sketch}. For every $(\vx,\vy)\!\in\!B_x$ (for example $p_1$ and $p_2$ in Figure \ref{sketch}), we construct a point $(\vx_r,\vy_s)$ which eventually bounds the trajectory starting from $(\vx,\vy)$; that is, we have $(\vx_r,\vy_s) \!\ll_K\! \phi_{t_1}(\vx,\vy) \!\leq_K\! (\vx^*,\0)$\footnote{$\phi_t(\vx,\vy)$ denotes the solution of \eqref{eq:biSIS} at $t\!\geq\!0$, with initial point $(\vx,\vy)$.} for some $t_1\!\geq\!0$. From the monotonicity shown in Proposition \ref{cooperative interor prop}, we have $\phi_t(\vx_r,\vy_s) \!\ll_K\! \phi_{t+t_1}(\vx,\vy) \!\leq_K\! (\vx^*,\0)$ for all time $t\!\geq\!0$. We prove that the trajectory starting from $(\vx_r,\vy_s)$ converges  to $(\vx^*,0)$ monotonically, with respect to the southeast cone-ordering (Figure \ref{sketch}(a)). Using this, we show the convergence of trajectories starting from $(\vx,\vy)$ via a sandwich argument (Figure \ref{sketch}(b)). See Appendix \ref{proof of the results} for detailed proof.
\end{IEEEproof}

\begin{figure*}[!t]
    \centering
    \captionsetup{justification=centering}
    \vspace{-4mm}
    \hspace{-0mm}
    \subfloat[{Virus 1 survives, Virus 2 dies: $\tau_1\lambda(\mS_{\vy^*}\mA)\!>\! 1$, $\tau_2 \lambda(\mS_{\vy^*}\mB)\!<\! 1$}]{\includegraphics[scale=0.22]{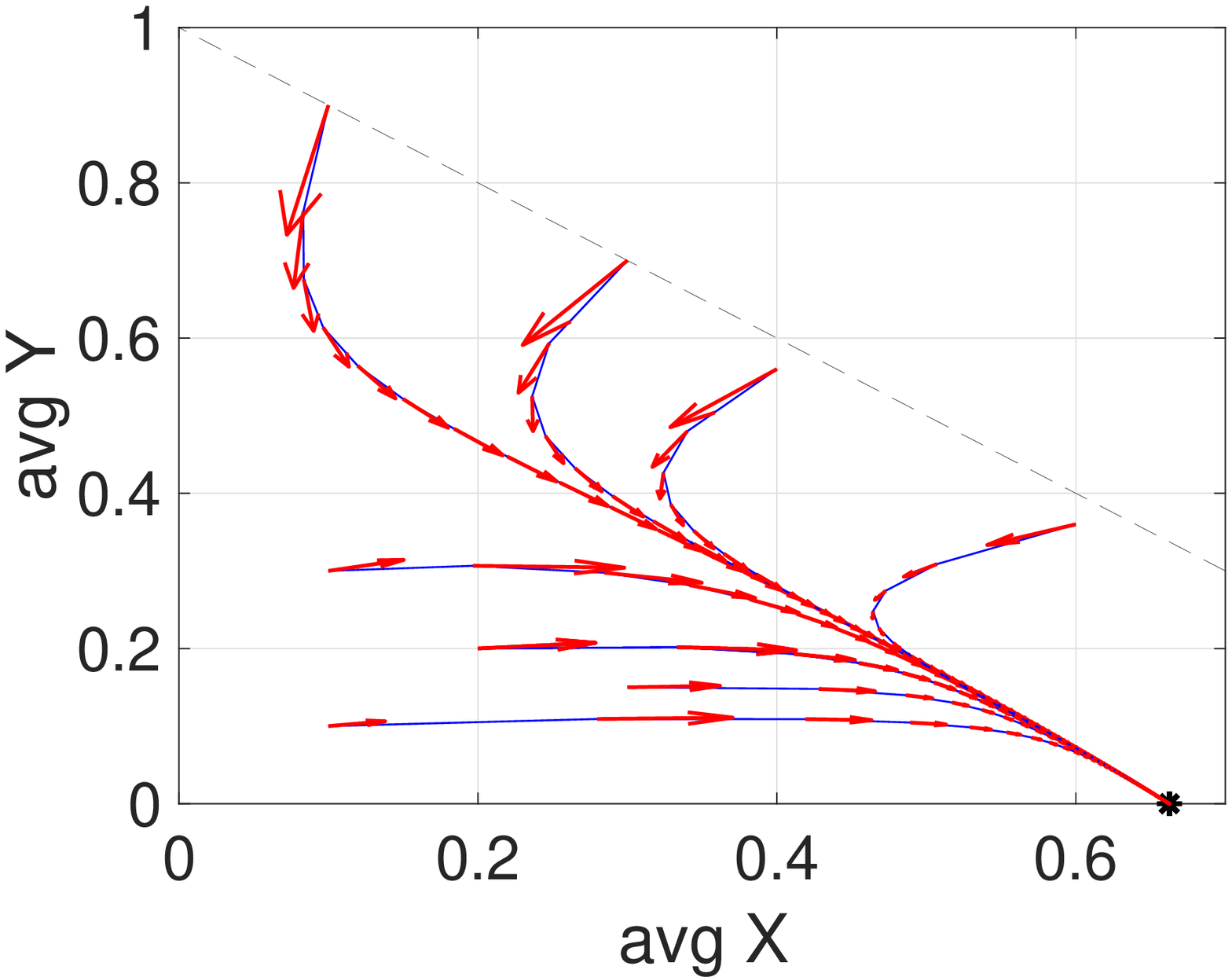}}
    \hfil
    \subfloat[Virus 2 survives, Virus 1 dies: $\tau_1 \lambda(\mS_{\vy^*}\mA)\!<\! 1$, $\tau_2 \lambda(\mS_{\vy^*}\mB) \!>\! 1$]{\includegraphics[scale=0.22]{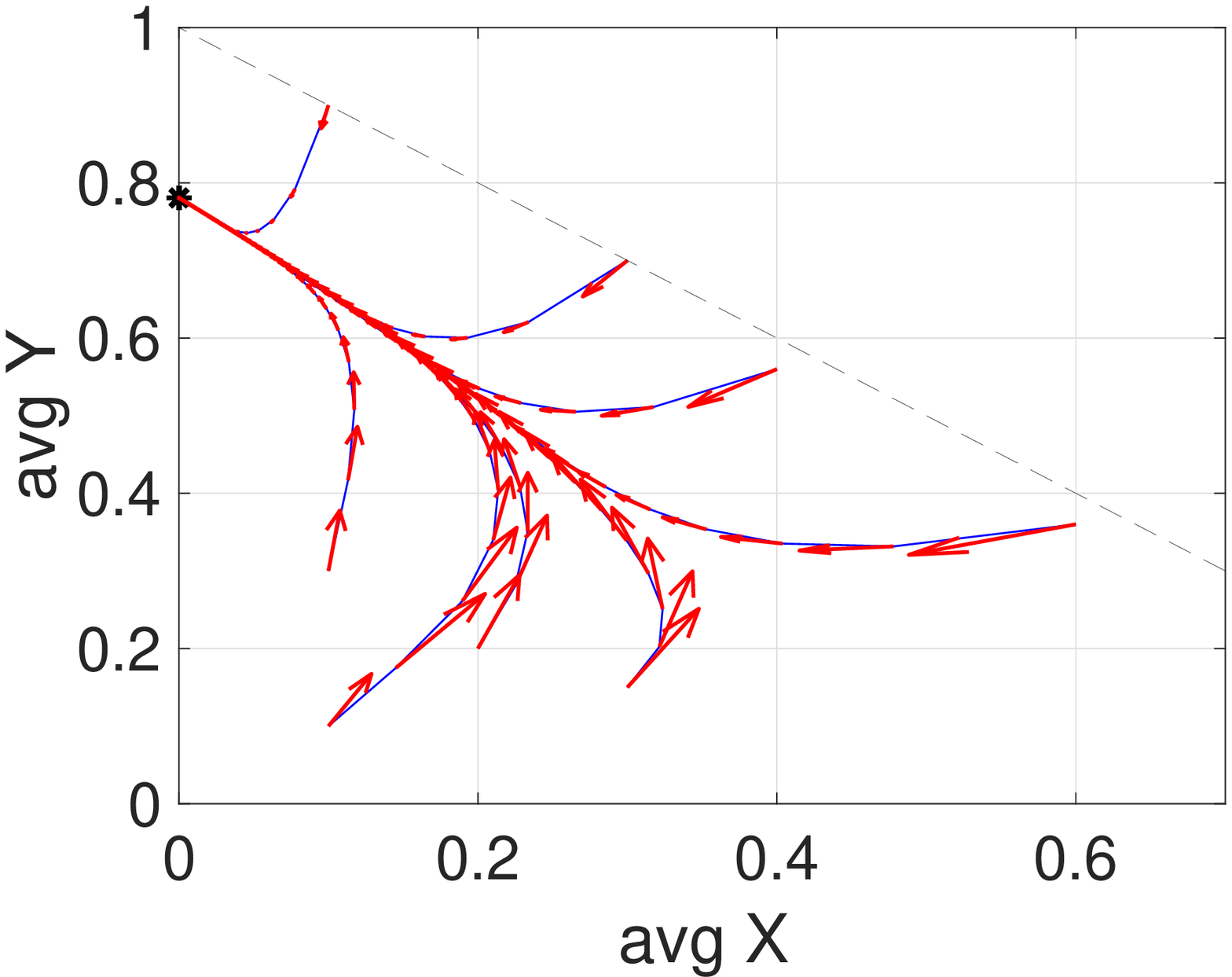}}%
    \label{fig:convergence}
    \hfil
    \subfloat[Coexistence: $\tau_1 \lambda(\mS_{\vy^*}\mA) \!>\! 1$, $\tau_2 \lambda(\mS_{\vy^*}\mB) \!>\! 1$]{\includegraphics[scale=0.22]{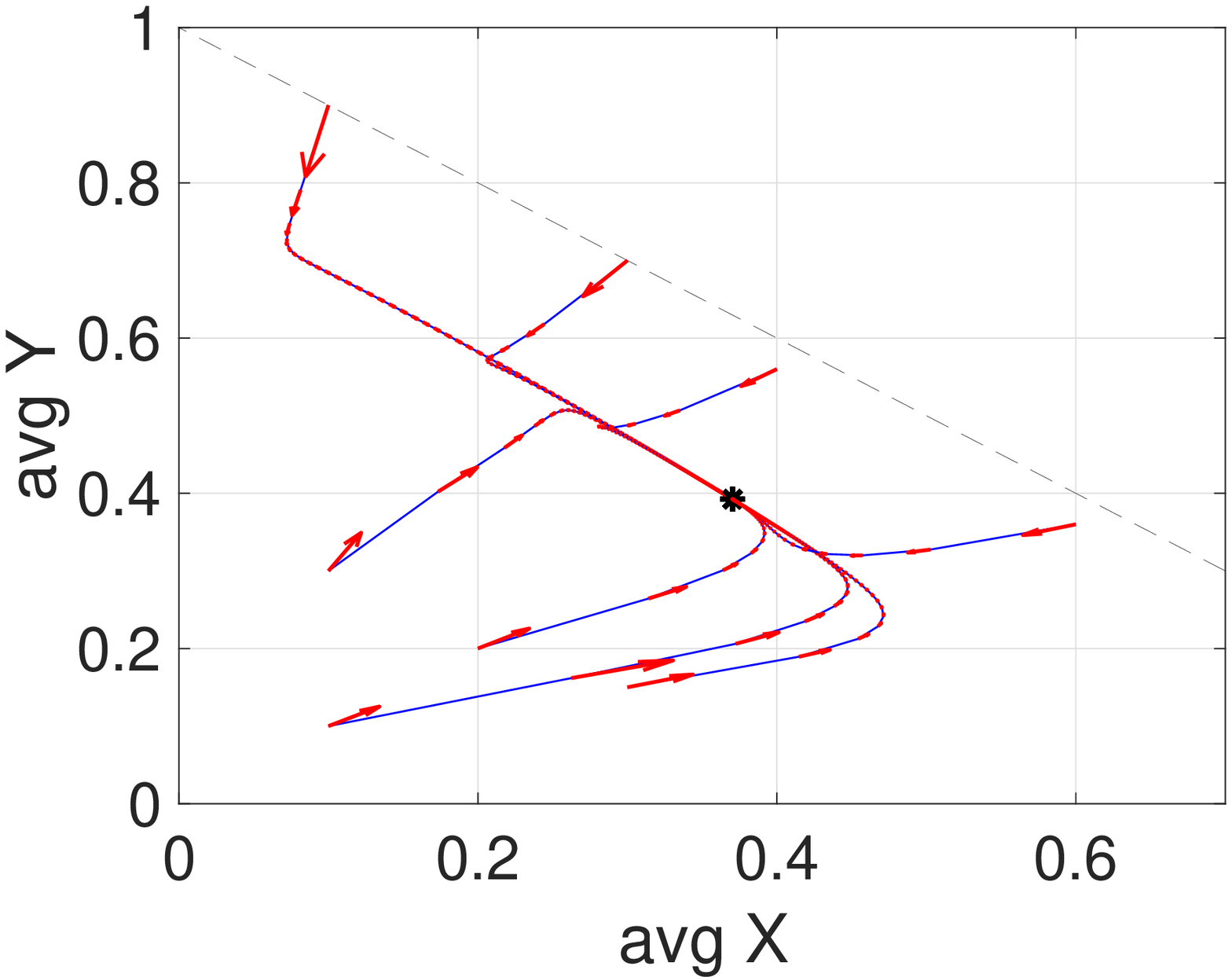} }
    \hfil
    \subfloat[Coexistence: $\tau_1 \lambda(\mS_{\vy^*}\mA) \!>\! 1$, $\tau_2 \lambda(\mS_{\vy^*}\mC) \!>\! 1$]{\includegraphics[scale=0.22]{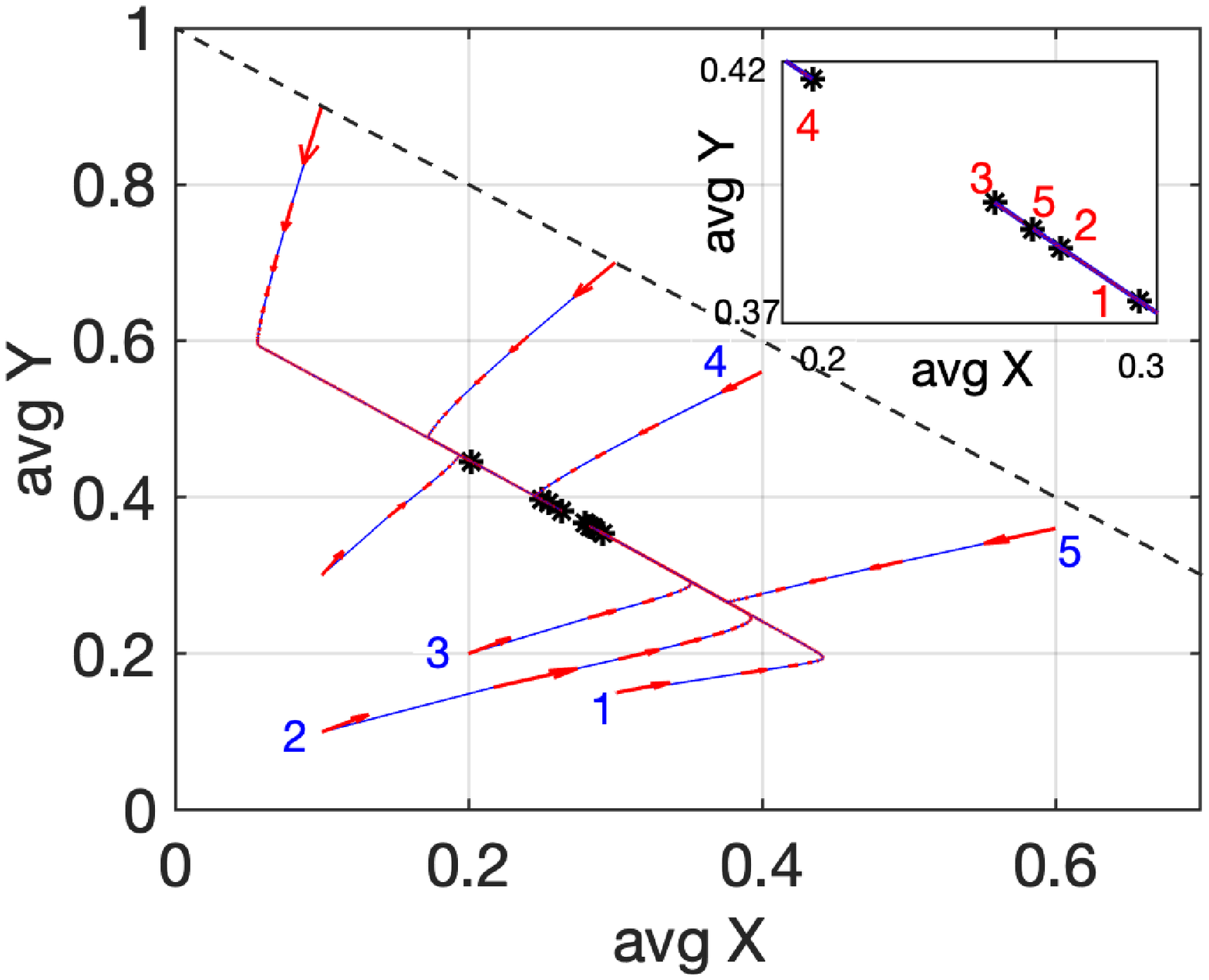} }

    \caption{Phase Plot results for simulation on the \textit{AS-733} graph.}\label{numerical figure}
    \vspace{-6mm}
\end{figure*}

From Theorem \ref{theorem wta}, we can deduce that the threshold values for $\tau_1$ and $\tau_2$ below which each of the viruses will die out are given by the equations $\tau_1\!=\!1/\lambda(\mS_{\vy^*} \mA )$ and $\tau_2\!=\!1/\lambda(\mS_{\vx^*} \mB )$, respectively. Figure \ref{all regions}(b) plots these threshold values for Virus 1 (in blue) and Virus 2 (in red) for varying values of $\tau_1$ and $\tau_2$, and partitions the entire parameter space into regions R1 -- R6 as shown. When $\tau_1 \!>\! 1/\lambda(\mA)$ and $\tau_2 \!>\!1/\lambda(\mB)$, for which values of $\tau_1,\tau_2$ do not lie in regions R1, R2 or R3, the blue curve lies above the red curve as in Figure \ref{all regions}(b). This was originally shown in \cite{sahneh2014competitive} by deducing that the ratio of slopes of the red and blue curves at point $(\tau_1,\tau_2) = \left(1/\lambda(\mA), 1/\lambda(\mB)\right)$ is less than one. This means there exist combinations of $\tau_1,\tau_2$ for which $\tau_1$ lies to the right of the blue curve ($\tau_1\lambda(\mS_{\vy^*} \mA ) \!>\! 1$), and $\tau_2$ lies above the red curve ($\tau_2\lambda(\mS_{\vx^*} \mB ) \!>\! 1$).\footnote{Note that $\tau_1\lambda(\mS_{\vy^*} \mA) \!\leq\! 1$ and $\tau_2\lambda(\mS_{\vx^*} \mB ) \!\leq\! 1$ is only possible in region R1, since it is the only region where $\tau_1$ can lie to the left of the blue curve, and $\tau_2$ can lie below the red curve. This effectively reduces the expressions to $\tau_1\lambda(\mA) \!\leq\! 1$ and $\tau_2\lambda(\mB ) \!\leq\! 1$, the conditions for convergence to the virus-free equilibrium as in Theorem \ref{theorem virus free}.} This corresponds to region R6 in Figure \ref{all regions}(b), and our final result shows that for values of $\tau_1,\tau_2$ which lie in R6, we observe convergence to coexistence equilibria. Let $E$ denote the set of all fixed points of the system in \eqref{eq:biSIS}.
\begin{theorem}[Convergence to coexistence equilibria]\label{theorem coexistence}
  If $\tau_1\lambda(\mS_{\vy^*} \mA) \!>\! 1$ and $\tau_2\lambda(\mS_{\vx^*} \mB ) \!>\! 1$, there exist fixed points $(\hat \vx, \hat \vy) \!\gg\! (\0,\0)$ and $(\bar \vx,\bar \vy) \!\gg\! (\0,\0)$ such that
  \begin{equation*}
      (\0,\vy^*) \ll_K (\hat \vx,\hat \vy) \leq_K (\bar \vx,\bar \vy) \ll_K (\vx^*,\0),
  \end{equation*}
  \noindent with the possibility that $(\hat \vx, \hat \vy) = (\bar \vx,\bar \vy)$. All trajectories starting from $B_x \cap B_y$ converge to the set of coexistence fixed points $S \triangleq \left\{ (\vx_e,\vy_e) \!\in\! E ~|~ (\hat \vx,\hat \vy) \!\leq_K\! (\vx_e,\vy_e) \!\leq_K\! (\bar \vx,\bar \vy)\right\}$.$\qedsymbol$
\end{theorem}

The proof of Theorem \ref{theorem coexistence} follows similar arguments to that of the previous theorem, and is the first convergence result for coexistence fixed points in the competing SIS literature. Note that while we have convergence to `a' coexistence equilibrium, it may or may not be unique in the state space. The global convergence is therefore to the set of possible coexistence equilibria, and not necessarily a singular point. Thus, via Theorems \ref{theorem virus free}, \ref{theorem wta} and \ref{theorem coexistence} we cover all possible convergence scenarios of the bi-virus SIS system \eqref{eq:biSIS}, and successfully establish the complete theoretical characterization for the trichotomy of possible outcomes.

\subsubsection*{\textbf{Comparison to existing literature}}
Now that we have established all our results, we briefly compare our work with results from \cite{yang2017bi,Santos2015}, which also talk about global convergence to single-virus equilibria. To this end, we illustrate the limitations of the existing conditions for global convergence in \cite{yang2017bi,Santos2015} in Figure \ref{all regions}(a); and use Figure \ref{all regions}(b), where we provide complete characterization of the parameter space, to draw comparisons with our results.

When translated to our setting, the result from \cite{Santos2015} says that when $\tau_1 d_{min}(\mA) \!>\! \tau_2 d_{max}(\mB)$, the Virus 2 is sure to die out (Virus 1 could persist or die out), and similarly when $\tau_1 d_{max}(\mA) \!<\! \tau_2 d_{min}(\mB)$, the Virus 1 is sure to die out. We illustrate these conditions in Figure \ref{all regions}(a), where Virus 1 (Virus 2) is sure to die out if parameters ($\tau_1,\tau_2$) lie above (below) the blue (red) line. Therefore, the entire yellow-shaded region in Figure \ref{all regions}(a), between the blue and red lines, is left uncharacterized in \cite{Santos2015}. When $\mA$ and $\mB$ are regular graphs with the same degree ($d_{min} \!=\! d_{max} \!=\! d$), the blue and red lines coincide, making coexistence infeasible. This is also mentioned in \cite{sahneh2014competitive} where they show that for regular graphs with same degree, the system behaves as if the two graphs were the same - rendering coexistence impossible (which is also in line with results in \cite{prakash2012winner}).
In contrast, the maximum degree of graphs can also be much larger than the minimum degree (e.g., power law graphs), causing the yellow-shaded space to become very large, possibly spanning almost the entire parameter space.

The main result in \cite{yang2017bi}, when similarly translated to our setting as above, says that when $\tau_1\lambda(\mA)\!>\!1$ and $\tau_2\lambda(\mB)\!\leq\!1$, Virus 1 survives and Virus 2 dies out. Similarly, when $\tau_2\lambda(\mB)\!>\!1$ and $\tau_1\lambda(\mA)\!\leq\!1$, Virus 2 survives and Virus 1 dies out. These correspond to regions R2 and R3 in Figure \ref{all regions}(b). However, their results do not cover the convergence properties for $\tau_1, \tau_2$ which lie in regions R4 -- R6. Our Theorems \ref{theorem wta} and \ref{theorem coexistence} do account for these values of $\tau_1, \tau_2$, and show convergence to $(\0,\vy^*)$, $(\vx^*,\0)$, or to a coexistence fixed point, whenever they lie in regions R4, R5, or R6, respectively.

In summary, without our Theorems \ref{theorem wta} and \ref{theorem coexistence},\footnote{All of our results can be extended to setting where the infection and recovery rates of the bi-virus model are \textit{heterogeneous}, as in \cite{liu2019analysis, Santos2015}. The adjacency matrices $\mA$ and $\mB$ can be symmetric, irreducible, weighted, with $a_{ij}, b_{ij} \!\geq\! 0$ (not necessarily 1/0-valued) multiplied by $\beta_1$ and $\beta_2$, respectively, being the infection rates from node $j\!\to\! i$ for Viruses 1 and 2. Recovery rates are similarly heterogenized as $\bfdelta_1=[\delta_1^i]$ and $\bfdelta_2=[\delta_2^i$] for viruses 1 and 2; written as recovery matrices $\text{diag}(\bfdelta_1)$ and $\text{diag}(\bfdelta_2)$, respectively.} results from literature fail to characterize a sizeable portion of the parameter space as shown in Figure \ref{all regions}(a) by the `\textbf{?}' region (part of the shaded region surrounded by the arrows). The parameters leading to coexistence are entirely contained in this region as well - explaining the dearth of convergence results for such equilibria in the existing literature.

\section{Numerical Results}\label{numerical results}

In this section, we present simulation results to support our theoretical findings. To this end, we consider an undirected, connected graph (103 nodes, 239 edges), called Autonomous System (AS-733), from the SNAP repository \cite{snapnets}. We generate two additional graphs, overlaid on the same set of nodes, by modifying the original graph (AS-733-A with $\lambda(\mA) \!=\! 12.16$), removing and adding edges while ensuring connectivity between the nodes. The graph AS-733-B has 716 edges with $\lambda(\mB) \!=\! 17.2$ while graph AS-733-C has 259 edges with $\lambda(\mC)\!\! =\!\! 12.26$. Note that since our theoretical results hold for any general graphs, we only use this set as example graphs only to numerically demonstrate the convergence properties. Similar numerical results can indeed be obtained for any other networks (such as social networks), and are omitted here due to space constraint.

We construct four cases that satisfy the assumptions of Theorems \ref{theorem wta} and \ref{theorem coexistence} which give global convergence to fixed points where (a) Virus 1 is the surviving epidemic (which spreads on graph AS-733-A), (b) Virus 2 is the surviving epidemic (which spreads on graph AS-733-B), (c) both viruses coexist with `a' fixed point, (where Virus 1 spreads on graph AS-733-A and Virus 2 on AS-733-B) and (d) both viruses coexist with multiple equilibrium points (where Virus 1 spreads on graph AS-733-A and Virus 2 on AS-733-C), respectively.
Table \ref{tab:parameters} summarizes these cases, with the parameter values therein satisfying the respective convergence conditions.

\begin{table}[h!]\vspace{-4mm}
\centering
\caption{Simulation Parameters}\vspace{-2mm}
\label{tab:parameters}
\resizebox{0.45\textwidth}{!}{%
\begin{tabular}{|c|c|c|}
\hline
\multicolumn{1}{|c|}{(a)} & {$\tau_1 \!=\! 1, ~\tau_2 \!=\! 0.15$} & $\lambda(\mS_{\vy^*}\!\mA) \!=\! 3.9, ~\lambda(\mS_{\vx^*}\!\mB) \!=\! 4.1$ \\ \hline
\multicolumn{1}{|c|}{(b)} & {$\tau_1 \!=\! 0.5, ~\tau_2 \!=\! 0.36$} & $\lambda(\mS_{\vy^*}\!\mA) \!=\! 1.5, ~\lambda(\mS_{\vx^*}\!\mB) \!=\! 6.4$   \\ \hline
\multicolumn{1}{|c|}{(c)} & {$\tau_1 \!=\! 1, ~\tau_2 \!=\! 0.36$} & $\lambda(\mS_{\vy^*}\!\mA) \!=\! 1.5, ~\lambda(\mS_{\vx^*}\!\mB) \!=\! 4.1$ \\ \hline
\multicolumn{1}{|c|}{(d)} & {$\tau_1 \!=\! 0.86, ~\tau_2 \!=\! 0.82$} & $\lambda(\mS_{\vy^*}\!\mA) \!=\! 1.19, ~\lambda(\mS_{\vx^*}\!\mC) \!=\! 1.25$ \\ \hline
\end{tabular}%
}
\vspace{-1mm}
\end{table}

In Figures \ref{numerical figure}(a) -- \ref{numerical figure}(d), we show numerical results for the above four convergence cases, respectively. To visualize our system in two dimensions, we use $avg X \!\triangleq\! (1/N)\sum_{i \in \cN} x_i$ on the x-axis, and $avg Y \!\triangleq\! (1/N)\sum_{i \in \cN} y_i$ on the y-axis. We plot trajectories of the bi-virus system starting from different initial points in the state space $D$ to observe their convergence, with red arrows representing the trajectories' direction of movement at various time intervals. Here, the state space $D$ is the region that lies below the dotted-line (in Figure \ref{numerical figure}), ensuring $x_i + y_i \leq 1$ for all $i \in \cN$, for every initial point.

Figures \ref{numerical figure}(a) and \ref{numerical figure}(b) show convergence to the two different single-virus equilibria. The values of $\tau_1,\tau_2$ for both of these cases satisfy the two sets of conditions in Theorem \ref{theorem wta}, respectively, thereby validating the result. Similarly, the parameters in cases (c) and (d) satisfy the coexistence conditions in Theorem \ref{theorem coexistence}, and we observe convergence to such equilibria as depicted in Figures \ref{numerical figure}(c) and (d). We observe a unique coexistence equilibrium when the viruses are competing over graphs AS-733-A and AS-733-B, for which the eigenvalues $\lambda(\mA)$ and $\lambda(\mB)$ are significantly different (Figure \ref{numerical figure}(c)). Interestingly, we also observe multiple coexistence equilibria, this time when the competition takes place over graphs AS-733-A and AS-733-C, for which the eigenvalues $\lambda(\mA)$ and $\lambda(\mC)$ are very close to each other. The `upper left' and `lower right' coexistence fixed points characterize the set $S$ of all such equilibria, as in Theorem \ref{theorem coexistence}. This can be observed more closely in the inset in Figure \ref{numerical figure}(d), where the number besides each fixed point (in red) corresponds to the different initial starting points (in blue) of the trajectories. Thus, convergence to set $S$ occurs globally over the state space, but exactly which coexistence fixed point the system converges to is dependent on the initial point. We are thus able to observe all possible convergence scenarios from Section \ref{convergence and coexistence}, including multiple coexistence equilibria.

\section{Concluding Remarks} \label{conclusion}
In this paper, we bridge the gap between linear stability properties and global convergence results (or lack thereof) for the bi-virus model in the literature, and succeed in providing a complete characterization of the trichotomy of possible outcomes for such competing epidemics - a well known open problem. We show that the bi-virus epidemic is monotone with respect to a specially constructed partial ordering, and draw techniques from monotone dynamical systems (MDS) to completely characterize the entire parameter space of the bi-virus system, a contrast to the usual Lyapunov based approach. Our results demonstrate how powerful these alternative proving techniques can be, compared to classical Lyapunov approaches; and we note that it may be worth exploring such monotonicity properties in other settings as well, where competition is the general theme.

\ifCLASSOPTIONcaptionsoff
  \newpage
\fi

\bibliographystyle{IEEEtran}
\bibliography{sis}

\appendices
\section{Basic Definitions and Results from Matrix Theory}\label{matrix theory results}
We first provide some well known results surrounding irreducible square matrices.

\begin{definition}\cite{meyer_textbook}\label{irreducible matrix}
  A square matrix $\mA$ is \textbf{reducible} if there exists a permutation matrix $\mP$ such that $\mP^T\mA\mP$ is a block diagonal matrix. If no such permutation matrix exists, we say that $\mA$ is \textbf{irreducible}.
\end{definition}
\noindent One way to check if a matrix is irreducible is by observing the underlying directed graph, where there is an edge between two nodes only if $a_{ij} \neq 0$. The matrix $A$ is irreducible if and only if this underlying directed graph is strongly connected.
\begin{definition}\cite{Berman1994book}\label{M-matrix}
  A \textbf{M-matrix} is a matrix with non-positive off-diagonal elements with eigenvalues whose real parts are non-negative.
\end{definition}

\noindent We use the following well known result for non-negative, irreducible matrices heavily throughout the paper.

\begin{theorem}(Perron-Frobenius)\label{PF theorem}\cite{meyer_textbook}
  Let $\mA$ be a non-negative, irreducible matrix. Then, $\lambda(\mA)$ is a strictly positive real number, and the corresponding eigenvector $\vecv$ where $\mA\vecv = \lambda(\mA)\vecv$ is also strictly positive. We call $\lambda(\mA)>0$ and $\vecv\gg \0$ the \textbf{PF eigenvalue} and \textbf{PF eigenvector} of the matrix respectively.$\qedsymbol$
\end{theorem}


\noindent The following result is on irreducible M-matrices.

\begin{lemma}\cite{Berman1994book}\label{M matrix lemma}
  Given an irreducible and non-singular M-matrix $\mM$, its inverse $\mM^{-1}$ has strictly positive entries.$\qedsymbol$
\end{lemma}

\section{Definitions and results from ODE literature}\label{ODE results}

We use the following definitions and results from the ODE literature throughout the paper.

\begin{definition}\label{flow definition}
  The `flow' of a dynamical system in a metric space $X$ is a map $\phi: X\!\times\! \R \!\to\! X$ such that for any $x_0 \!\in\! X$ and all $s,t \in \R$, we have $\phi_0(x_0) \!=\! x_0$ and $\phi_s\left( \phi_t(x_0) \right) \!=\! \phi_{t+s}(x_0)$.
\end{definition}

\begin{definition}
  A flow $\phi:X\times \R \to X$ is \textbf{positively invariant} in set $P\subset X$ if for every $x_0 \in P$, $\phi_t(x_0) \in P$ for all $t > 0$.
\end{definition}

\begin{definition}\label{fixed point definiton}
  Given a flow $\phi$, an `equilibrium' or a `fixed point' of the system is a point $x^* \in X$ such that $\{x^*\}$ is a positively invariant set. For the ODE system
  $\dot x = F(x)$,
  we have $F(x^*) = 0$ at the equilibrium.
\end{definition}

\noindent For an equilibrium point $x^*\in X$ we say that the trajectory starting at $x_0\in X$ \textit{converges} to $x^*$ if $\lim_{t \to \infty} \phi_t(x_0) = x^*$. The following result is true for stable fixed points of the ODE system from Definition \ref{flow definition}.

\begin{proposition}\cite{Perko2001} \label{Manifold}
  Let $\mJ F(x_0)$ be the Jacobian of the ODE system evaluated at a fixed point $x_0$ and assume it to be an irreducible matrix. Let $\lambda\left( \mJ F(x_0) \right) < 0$ and suppose the corresponding eigenvalue $\vecv$ is strictly positive ($\vecv \gg \0$). Then, there exists an $\epsilon > 0$ such that $F(x_0 + r\vecv) \ll 0$ for all $r \in (0,\epsilon]$ and $F(x_0 + r\vecv) \gg 0$ for all $r \in (0,-\epsilon]$\footnotemark.$\qedsymbol$
\end{proposition}

\footnotetext{In other words eigenvector $\vecv$ is tangent to the stable manifold of the ODE system at the stable fixed points $x_0$.}

\section{Results from MDS and Cooperative Systems}\label{MDS}

\begin{definition}\cite{Hirsch-I, HLSmith'88,HLSmith'17}\label{monotone and co-operative}
  A flow $\phi$ is said to be \textbf{monotone} if for all $\vx, \vy \in \R^n$ such that $\vx \leq_K \vy$ and any $t\geq0$, we have
  $\phi_t(\vx) \leq_K \phi_t(\vy).$
  
  If the flow represents the solution of an ODE system, we say that the ODE system is \textbf{co-operative}.
\end{definition}

\begin{definition}\label{irreducible ode}
  Consider the system \eqref{ode sys} and let $\mJ F(\vx)\! \triangleq\! \left[{df_i(\vx)}/{dx_j}\right]$ be the Jacobian of the right hand side evaluated at any point $\vx \! \in \! \R^n$. We say that \eqref{ode sys} is an \textbf{irreducible ODE} in set $D \in \R^n$ if for all $\vx \in D$, $\mJ F(\vx)$ is an irreducible matrix.
\end{definition}
\begin{definition}\cite{HLSmith'88,MDSbook,HLSmith'17}\label{strongly monotone}
  The flow $\phi$ is said to be \textbf{strongly monotone} if it is monotone, and for all $\vx,\vy \in \R^n$ such that $\vx <_K \vy$, and time $t \geq 0$, we have
  $\phi_t(\vx) \ll_k \phi_t(\vy).$
\end{definition}
\begin{theorem}\label{SM ODE}\cite{HLSmith'88,MDSbook,HLSmith'17}
  Let \eqref{ode sys} be irreducible and co-operative in some set $D \subset \R^n$. Then the solution $\phi$ (restricted to $t \geq 0$) is strongly monotone.$\qedsymbol$
\end{theorem}
As part of the main result of monotone dynamical systems, trajectories of strongly monotone systems, starting from almost anywhere (in the measure theoretic sense) in the state space, converge to the set of equilibrium points \cite{Hirsch-II,HLSmith'91,HLSmith'04,HLSmith'17}. However, often the systems are strongly monotone only in the interior of the state spaces instead of the entirety of the state space. In such cases, the following results are useful.

\begin{proposition}(Proposition 3.2.1 in \cite{MDSbook})\label{invariance prop}
Consider the ODE system \eqref{ode sys} which is cooperative in a compact set $D\subset \R^n$ with respect to some cone-ordering, and let $<_r$ stand for any of the order relations $\leq_K, <_K, \ll_K$. Then, $P_+ \!\triangleq\! \left\lbrace \vx \!\in\! D ~|~ \0 \!<_r\! F(\vx) \right\rbrace$ and $P_- \!\triangleq\! \left\lbrace \vx \!\in\! D ~|~ F(\vx) \!<_r\! \0 \right\rbrace$ are positively invariant, and the trajectory $\left\lbrace \phi_t(\vx) \right\rbrace_{t \geq 0}$ for any point $\vx \!\in\! P_+$ or $\vx \!\in\! P_-$ converges to an equilibrium.$\qedsymbol$
\end{proposition}

\begin{theorem}(Theorem 4.3.3 in \cite{MDSbook})\label{existence of another fp}
Let \eqref{ode sys} be cooperative (with respect to some cone-ordering $\leq_K$) in a compact set $D \subset \R^n$ and let $\vx_0 \in D$ be an equilibrium point. Suppose that $s \triangleq \lambda(\mJ F(\vx_0))>0$ (i.e. $\vx_0$ is an unstable fixed point) and there is an eigenvector $\vecv \gg_K \0$ such that $\mJ F(\vx_0) \vecv = s \vecv$. Then, there exists $\epsilon_0 \in (0,\epsilon]$ and another equilibrium point $\vx_e$ such that for each $r \in (0,\epsilon_0]$, the solution $\phi_t(\vx_r)$ has the following properties:
\begin{itemize}
\item[(1)]
$\vx_r \!\ll_K\! \phi_{t_1}(\vx_r)\! \ll_K\! \phi_{t_2}(\vx_r)\! \ll_K \!\vx_e$, for any $0\!<\!t_1\!<\!t_2$.
\item[(2)]
${d \phi_t(\vx_r)}/{dt} \gg_K \0$, for any $t>0$.
\item[(3)]
$\phi_t(\vx_r) \rightarrow \vx_e$, as $t \rightarrow \infty$.$\qedsymbol$
\end{itemize}
\end{theorem}

\section{Proofs of the Main Results}\label{proof of the results}

Throughout this Appendix, we use $\phi_t(\vx_0,\vy_0)$ to represent the solution of \eqref{eq:biSIS} at time $t \geq 0$, starting from $(\vx_0,\vy_0) \in D$. We will need the following results to prove the theorems from Section \ref{convergence and coexistence}.

\begin{proposition}\label{convergence to Z}
  Starting from any point $D \setminus \left\{ (\0,\0) \right\}$, trajectories of \eqref{eq:biSIS} converge to the set
  \begin{equation*}
      ~~~~Z \triangleq \left\{ (\vu,\vw) \in D ~|~ (\0,\vy^*)\leq_K (\vu,\vw) \leq_K (\vx^*, \0) \right\}.~~~\qedsymbol
  \end{equation*}
\end{proposition}
\begin{IEEEproof}
  For any $(\vr,\vs) \in D\setminus \{(\0,\0)\}$, there exists points $\vx,\vy \in [0,1]^N$ such that $(\0,\vy) \leq_K (\vr,\vs) \leq_K (\vx,\0)$. Then, from Definition \ref{monotone and co-operative} of a monotone system, we have $\phi_t(\0,\vy) \leq_K \phi_t(\vr,\vs) \leq_K \phi_t(\vx,\0)$ for any $t > 0$. Since $\phi_t(\vx,\0)\rightarrow(\vx^*,\0)$ and $\phi_t(\0,\vy) \rightarrow (\0,\vy^*)$, we get $(\0,\vy^*) \leq_K \lim_{t \rightarrow \infty} \phi_t(\vr,\vs) \leq_K  (\vx^*,\0)$. Thus the trajectory $\left\{\phi_t(\vr,\vs)\right\}_{t \geq 0}$ converges to $Z$, completing the proof.
\end{IEEEproof}

Since the set $Z$ depends on $\vx^*$ and $\vy^*$, the fixed points of systems \eqref{sis-x} and \eqref{sis-y}, and we can determine when these fixed points are positive or zero, Proposition \ref{convergence to Z} helps us to quickly point out a subset of the state space to which trajectories starting from any point in $D\!\setminus\! \left\{ (\0,\0) \right\}$ converge.

\begin{IEEEproof}[\textbf{Proof of Theorem \ref{theorem virus free}}]
  When $\tau_1\lambda(\mA) \leq 1 ~\text{and}~ \tau_2\lambda(\mB) \leq 1$, we know from Theorem \ref{sis-conditions} that $\vx^* = \vy^* = 0$. Therefore, trajectories of \eqref{eq:biSIS} starting from any point in $D\setminus \left\{ (\0,\0) \right\}$ converge to the set $Z \triangleq \left\{ (\vu,\vw) \in D ~|~ (\0,\0)\leq_K (\vu,\vw) \leq_K (\0, \0) \right\} = \left\{ (\0,\0) \right\}$. Hence, the virus-free equilibrium is globally asymptotically stable in $D$, which completes the proof.
\end{IEEEproof}

Proposition \ref{convergence to Z} can also be applied to show that $(\vx^*,\0)$ where $\vx^* \gg \0$ is globally attractive when $\tau_1\lambda(\mA) > 1$ and $\tau_2\lambda(\mB) \leq 1$. This is because from Theorem \ref{sis-conditions}, we know that $\vx^* \gg \0$ and $\vy^*=\0$. We then have $Z \triangleq \left\{ (\vu,\vw) \in D ~|~ (\0,\0)\leq_K (\vu,\vw) \leq_K (\vx^*, \0) \right\}$, implying that the system \eqref{eq:biSIS} ultimately reduces to the single $SIS$ system \eqref{sis-x}, which we know globally converges to $\vx^*$. By a symmetric argument, we also have that $(\0,\vy^*)$ where $\vy^* \gg \0$ is globally attractive when $\tau_1\lambda(\mA) \leq 1$ and $\tau_2\lambda(\mB) > 1$. Therefore these cases are easily analyzed by applying Proposition \ref{convergence to Z} in conjunction with Theorem \ref{sis-conditions}. The values of $\tau_1$ and $\tau_2$ which satisfy these conditions, lie in regions R2 and R3 of Figure \ref{all regions}(b) and we henceforth exclude them from our analysis, considering only those values of $\tau_1$ and $\tau_2$ for which $\tau_1\lambda(\mA)\!>\!1$ and $\tau_2\lambda(\mB)\!>\!1$ always holds. Thus, $\vx^*$ and $\vy^*$ are henceforth implied to be strictly positive vectors.

Before formally proving Theorems \ref{theorem wta} and \ref{theorem coexistence}, we provide some additional constructions and notations which will help simplify the proofs. As in Remark \ref{remark sis coop}, the Jacobians $\mJ F^x(\vx)$ and $\mJ F^y(\vy)$ of systems \eqref{sis-x} and \eqref{sis-y}, respectively, are\vspace{-1mm}
\begin{align*}\vspace{-1mm}
  \mJ F^x(\vx) &\!=\! \beta_1\text{diag}(\ones \!\!-\! \vx)\mA \!-\! \beta_1\text{diag}(\mA \vx) \!-\! \delta_1 \eye, ~~\forall \vx \in [0,\!1]^N, \\
  \mJ F^y(\vy) &\!=\! \beta_2\text{diag}(\ones \!\!-\! \vy)\mA \!-\! \beta_2\text{diag}(\mB \vy) \!-\! \delta_2 \eye, ~~\forall \vy \in [0,\!1]^N.
\end{align*}
\noindent Now recall the Jacobian $\mJ_{GH}(\vx,\vy)$ of the bi-virus ODE \eqref{eq:biSIS} from \eqref{Jacobian bi-virus}. When evaluated at $(\vx^*,\0)$ and at $(\0,\vy^*)$, we get\vspace{-1.5mm}
\begin{equation}\label{Jacobian wta x}\vspace{-1.5mm}
  \begin{split}
    \mJ_{GH}(\vx^*,\0)=
    \begin{bmatrix}
    \mJ F^x(\vx^*)      & \mP \\
         \0           & \mJ_y
    \end{bmatrix}
  \end{split}
\end{equation}
\noindent where $\mP \!=\! -\beta_1\text{diag}(\mA\vx^*)$, $\mJ_y \!=\! \beta_2\text{diag}(\ones \!-\! \vx^*)\mB \!-\! \delta_2 \eye$, and\vspace{-1.5mm}
\begin{equation}\label{Jacobian wta y}\vspace{-1.5mm}
  \begin{split}
    \mJ_{GH}(\0,\vy^*)=
    \begin{bmatrix}
    \mJ_x          & \0 \\
     \mQ       & \mJ F^y(\vy^*)
    \end{bmatrix}
  \end{split}
\end{equation}
\noindent where $\mQ \!=\! -\beta_2\text{diag}(\mB\vy^*)$, $\mJ_x \!=\! \beta_1\text{diag}(\ones \!-\! \vy^*)\mA \!-\! \delta_1 \eye$. This leads us to the following proposition, where the ordering $\leq_K$ ($<_K,\ll_K$) stands for the south east cone-ordering.

\begin{proposition}\label{positive ev}
  When $\tau_1\lambda(\mS_{\vy^*}\mA)\!>\!1$, we have $\lambda\left( \mJ_{GH}(0,\vy^*) \right) \!=\! \lambda(\mJ_x) \!>\! 0$, and the corresponding eigenvector $(\vu,\vecv) \!\in\! \R^{2N}$ of $\mJ_{GH}(0,\vy^*)$ satisfies $(\vu,\vecv) \!\gg_K\! (\0,\0)$.$\qedsymbol$
\end{proposition}
\begin{IEEEproof}
  First, recall that $\vy^* \gg \0$ is the asymptotically stable fixed point of \eqref{sis-y}. This implies that the real parts of all eigenvalues of the Jacobian  $\mJ F^y(\vy^*)$ of \eqref{sis-y} evaluated at $\vy^*$ are negative. Since $\mJ F^y(\vy^*)$ is an irreducible matrix as discussed in Section \ref{monotonicity bi-virus}, with non-negative off-diagonal elements, its PF eigenvalue (obtained by perturbing with a large multiple of the identity matrix) is real and negative, that is $\lambda\left(\mJ F^y (\vy^*)\right)<0$.

  Observe that $\tau_1\lambda(\mS_{\vy^*}\mA) \!>\! 1 $ implies that $\lambda(J_x) \!=\! \lambda(\beta_1\mS_{\vy^*}\mA \!-\! \delta_1 \eye) \!>\! 0$. Since $\mJ_{GH}(\0,\vy^*)$ is a block triangle matrix, we have $\lambda\left(\mJ_{GH}(\0,\vy^*)\right) \!=\! \max\! \left\{ \lambda(J_x), \lambda\left(\mJ F^y (\vy^*)\right) \right\}$, and since $\lambda\left(\mJ F^y (\vy^*)\right) \!<\! 0$, we obtain $\lambda\left(\mJ_{GH}(\0,\vy^*)\right) \!=\! \lambda(J_x) \!>\! 0$. Then, the corresponding eigenvector $(\vu,\vecv)$ satisfies\vspace{-1mm}
  \begin{equation*}\vspace{-1mm}
    \mJ_x \vu \!=\! \lambda(\mJ_x)\vu ~~~~\text{and}~~~~
    \mQ\vu \!+\! \mJ F^y(\vy^*)\vecv \!=\! \lambda(\mJ_x)\vecv.
  \end{equation*}
  \noindent From the first equation, we can tell that $\vu$ is the eigenvector of $\mJ_x$ corresponding to its PF eigenvalue, and thus satisfies $\vu \!\gg\! \0$. Now recall that $\mJ F^y(\vy^*)$ had eigenvalues with strictly negative real parts. $ \lambda(\mJ_x)\eye \!-\!\mJ F^y(\vy^*)$ is then a matrix with eigenvalues having strictly positive real parts (since $\lambda(\mJ_x)\!>\!0$). The matrix $\mM \triangleq \lambda(\mJ_x) \!-\! \mJ F^y(\vy^*)$ is then, by Definition \ref{M-matrix}, an M-matrix. By construction, it is also irreducible and invertible and from Lemma \ref{M matrix lemma}, we obtain that $\mM^{-1}$ is a (strictly) positive matrix. The second equation in the above can then be rewritten as $\vecv = \mM^{-1}\mQ\vu \ll \0$, where the inequality is because $\mQ \!=\! -\beta_2 \text{diag}(\mB\vy^*)$ has strictly negative diagonal elements. Therefore, since $\vu \gg \0$ and $\vecv \ll \0$, we have $(\vu,\vecv) \gg_K \0$, completing the proof.
\end{IEEEproof}

The intention behind introducing Proposition \ref{positive ev} was to satisfy the assumptions of Theorem \ref{existence of another fp}. In particular, when $\tau_1\lambda(\mS_{\vy^*}\mA)>1$, $(0,\vy^*)$ is an unstable fixed point, by Proposition \ref{positive ev} and Theorem \ref{existence of another fp}, there exists an $\epsilon_1 > 0$ and another fixed point $(\vx_e,\vy_e)$ such that for any point $(\vx_r,\vy_r) \triangleq (\0,\vy^*) + r(\vu,\vecv)$ where $r \in (0,\epsilon_1]$, we have \vspace{-3mm}
\begin{equation*}\label{monotone seq to xe,ye}\vspace{-2mm}
  \begin{split}
    (0,\vy^*) \!\ll\! (\vx_r,\vy_r) \!\ll_K\! \phi_t(\vx_r,\vy_r)
                                \!\ll_K\! \phi_s(\vx_r,\vy_r) \!\leq_K\! (\vx^*,\0)
  \end{split}
\end{equation*}
\noindent for all $s \!>\! t \!>\! 0$. Moreover, for all $(\vx,\vy)$ such that $(\0,\vy^*) \!\ll_K\! (\vx,\vy) \!\leq_K\! (\vx_e,\vy_e)$, there exists an $r\!\in\!(0,\epsilon]$ sufficiently small such that $(\vx_r,\vy_r) \!\leq_K\! (\vx,\vy) \!\leq_K\! (\vx_e,\vy_e)$. Since $\phi_t(\vx_r,\vy_r) \!\rightarrow\! (\vx_e,\vy_e)$, monotonicity implies $\phi_t(\vx,\vy) \!\rightarrow\! (\vx_e,\vy_e)$ as $t\!\to\!\infty$.

Now, we can either have $(\vx_e,\vy_e) \!=\! (\vx^*,\0)$, which occurs when $(\vx^*,\0)$ is the other stable fixed point of \eqref{eq:biSIS}, or $(\vx_e,\vy_e) \!=\! (\hat \vx, \hat \vy) \!\gg\! \0$ which occurs when $(\vx^*,\0)$ is an unstable fixed point. Note that  $(\vx^*,\0)$ is stable (unstable) if and only if $\tau_1\lambda(\mS_{\vy^*} \mA ) \!\leq\! 1$ ($ >\!\! 1$). We will talk about both these possibilities one by one and exploring these will eventually lead to Theorems \ref{theorem wta} and \ref{theorem coexistence}. But before we do that, we first prove the following proposition about convergence to the fixed point $(\vx_e,\vy_e)$ (whichever of the two it may be).

\begin{proposition}\label{convergence to x_e,y_e}
  Trajectories of the system \eqref{eq:biSIS} starting from any point $(\vx,\vy)$ such that $(\0,\vy^*) <_K (\vx,\vy) \leq_K (\vx_e,\vy_e)$ converge to $(\vx_e,\vy_e)$.$\qedsymbol$
\end{proposition}
\begin{IEEEproof}
  Recall that we already know that for all $(\0,\vy^*) \!\ll_K\! (\vx,\vy) \!\leq\! (\vx^*,\vy^*)$, $\phi_t(\vx,\vy) \!\rightarrow\! (\vx^*,\0)$. We would however like to show this for all $(\vx,\vy)\in Z\setminus(\0,\vy^*)$, that is even when $(\vx,\vy)$ satisfies $(\0,\vy^*) <_K (\vx,\vy) \leq (\vx^*,\0)$. To do this, we create a set of points which converge to $(\vx^*,\0)$, just like we created $(\vx_r,\vy_r)$ before, and then use a monotonicity argument to show convergence to $(\0,\vy^*)$ of trajectories starting for all points $(\vx,\vy)$ satisfying $(\0,\vy^*) <_K (\vx,\vy) \leq (\vx^*,\0)$.

  Recall that, $\vy^*$ is an asymptotically stable fixed point of \eqref{sis-y}, and from the proof of Proposition \ref{positive ev} we know that $\lambda\left(\mJ F^y (\vy^*)\right)<0$. Let $\vw \gg \0$ be the corresponding PF eigenvector. Then by Proposition \ref{Manifold}, there exists an $\epsilon_2 > 0$ such that for all $s \in (0,\epsilon_2]$, $F^y(\vy^* + s \vw) \ll \0$. We can then define points $(\vx_r,\vy_s) \triangleq (r*\vu,\vy^* + s\vw)$ for any $r \in (0,\epsilon_1]$ and $s \in (0,\epsilon_2]$, where $\vu\gg \0$ is the eigenvector of $\mJ_x$ from Proposition \ref{positive ev}. We will first show that trajectories starting from these points converge to $(\vx_e,\vy_e)$. By rearranging the terms of \eqref{eq:biSIS}, we can rewrite it as
  \begin{align*}
    \dot \vx &= \beta_1\text{diag}(\ones - \vy^*)\mA\vx - \delta_1\vx + \beta_1\text{diag}(\vy^*-\vx-\vy)\mA\vx \\
             &= \mJ_x\vx + O\left(\| \vx \|\left[\| \vy - \vy^* \| + \| \vx \|\right] \right), \\
    \dot \vy &= \beta_2\text{diag}(\ones - \vy)\mB\vy - \delta_2\vy - \beta_2\text{diag}(\vx)\mB\vy \\
             &= F^y(\vy) + O\left(\| \vy \|\right),
  \end{align*}
  \noindent for all $(\vx,\vy) \in D$\footnotemark. For any point $(\vx_r,\vy_s) = (r*\vu,\vy^* + s\vw)$, the above equations can be written as
  \footnotetext{Here, $O(x)$ is used to represent terms which satisfy $O(x) \to 0$ as $x \to 0$.}
  \begin{align*}
    \dot \vx &\!\!=\!\! r \lambda(\mJ_x) \vu \!+\! rO\left(\| \vu \|\left[s\| \vw \| \!+\! r\| \vu \|\right] \right)\!=\! r \!\left[\lambda(\mJ_x)\vu \!+\! O(r\!+\!s) \!\right]\!, \\
    \dot \vy &\!=\! F^y(\vy^* \!+\! s\vw) \!+\! O\left(\| s \vy \|\right) = F^y(\vy^* \!+\! s\vw) \!+\! O\left( s \right).
  \end{align*}
  \noindent For sufficiently small $r$ and $s$, we have $\dot \vx \gg \0$ (since $\lambda(\mJ_x) >0$ and $\vu \gg \0$) and $\dot \vy \ll \0$ (since $F^y(\vy^* + s \vw) \ll \0$ for all $s\in(0,\epsilon_2]$). This satisfies the conditions for Proposition \ref{invariance prop}, and trajectories starting from such points will be monotonically increasing (according to the south-east cone ordering), eventually converging to the fixed point $(\vx_e,\vy_e)$.

  Now see that for any point $(\vx,\vy)$ such that $(\0,\vy^*) \!<_K\! (\vx,\vy) \!\leq_K\! (\vx_e,\vy_e)$, where $\vx \!>\! \0$ and $\vy \!\leq\! \vy^*$, by the nature of the ODE system \eqref{eq:biSIS} all zero entries of the $\vx$ term will eventually become positive (if it isn't already). Therefore, there exists a time $t_1>0$ such that $\vx(t_1) \!\gg\! \0$, and there exist $r,s$ small enough such that $(\vx_r,\vy_s) \!\ll_K\! \phi_{t_1}(\vx,\vy) \!\leq_K\! (\vx_e,\vy_e)$. Again by monotonicity, since $\phi_t(\vx_r,\vy_s) \rightarrow (\vx_e,\vy_e)$, we have $\phi_{t+t_1}(\vx,\vy)\rightarrow(\vx_e,\vy_e)$ as $t \rightarrow \infty$, completing the proof.
\end{IEEEproof}
We now consider the case where $(\vx_e,\vy_e) \!=\! (\vx^*, \0)$ and give the proof for Theorem \ref{theorem wta}. We prove it only for when $\tau_1\lambda(\mS_{\vy^*} \mA ) \!>\! 1$ and $\tau_2\lambda(\mS_{\vx^*} \mB) \!\leq\! 1$, since the other case follows by a symmetric argument.

\begin{IEEEproof}[\textbf{Proof of Theorem \ref{theorem wta}}]
  When $\tau_2\lambda(\mS_{\vx^*} \mB) \leq 1$, $(\vx^*,\0)$ is a stable fixed point of system \eqref{eq:biSIS}, since all eigenvalues of $\mJ_{GH}(\vx^*,\0)$ have non-positive real parts, and we have $(\vx_e,\vy_e) = (\vx^*,\0)$. Proposition \ref{convergence to x_e,y_e} then implies that trajectories starting from all points in $Z\setminus \left\{ (\0,\vy^*) \right\}$ converge to $(\vx^*,\0)$. According to Proposition \ref{convergence to Z}, trajectories starting from all points $(\vx,\vy)\in B_x$ in the system eventually enter the set $Z$, thereby eventually converging to $(\vx^*,\0)$, giving us global convergence in $B_x$.
\end{IEEEproof}

Similarly, we use Propositon \ref{convergence to x_e,y_e} to prove Theorem \ref{theorem coexistence}.

\begin{IEEEproof}[\textbf{Proof of Theorem \ref{theorem coexistence}}]
  When $\tau_1\lambda(\mS_{\vy^*} \mA ) \!>\! 1$ and $\tau_2\lambda(\mS_{\vx^*} \mB) \!>\! 1$, both $(\0,\vy^*)$ and $(\vx^*,\0)$ are unstable fixed points, and $(\vx_e,\vy_e)$ takes the form of a positive fixed point $(\hat \vx, \hat \vy) \gg \0$ (it cannot be $(\vx^*,\0)$, which is unstable). Then from Proposition \ref{convergence to x_e,y_e}, it attracts trajectories beginning from all points $(\vx,\vy)$ satisfying $(\0,\vy^*) \!<_K\! (\vx,\vy) \!\leq_K\! (\hat \vx,\hat \vy)$.

  Similarly, we have a symmetric result beginning from $\tau_2\lambda(\mS_{\vx^*} \mB) \!>\! 1$ (symmetric to Proposition \ref{positive ev} which assumes $\tau_1\lambda(\mS_{\vy^*} \mA ) \!>\! 1$ instead), and we can say that there exists another fixed point $(\bar \vx, \bar \vy) \!\gg\! \0$ which attracts all points $(\vx,\vy)$ satisfying $(\bar \vx, \bar \vy) \!\leq_K\! (\vx,\vy)\!<_K\!(\vx^*,\0)$. By construction, we then have $(\hat \vx , \hat \vy) \!\leq_K\! (\bar \vx, \bar \vy)$, with the possibility of being equal.

  To prove global convergence of the system to the set $S \!=\! \left\{ (\vx_e,\vy_e) \!\in\! E ~|~ (\hat \vx,\hat \vx) \!\leq_K\! (\vx_e,\vy_e) \!\leq_K\! (\bar \vx,\bar \vy)\right\}$, observe first that as part of the proof of Proposition \ref{convergence to x_e,y_e} we showed that for trajectories starting from any point $(\vx,\vy)$ in the state space, there exists $r\!>\!0$ and $s\!>\!0$ small enough, and $t_1\!>\!0$ such that $(\vx_r,\vy_s) \!\ll_K\! \phi_{t_1}(\vx,\vy) \!\leq_K\! (\hat \vx,\hat \vy)$ where $(\vx_r,\vy_s)$ is a point very close to $(\vx^*,\0)$. By a parallel argument, we can find a similar point $(\vx_p,\vy_q)$ very close to $(\0,\vy^*)$ and a time $t_2$ such that $(\bar \vx,\bar \vy) \!\leq_K\! \phi_{t_2}(\vx,\vy) \!\ll_K\! (\vx_p,\vy_q)$. Then, we have $(\vx_r,\vy_s) \!\ll_K\! \phi_{\max\{t_1,t_2\}}(\vx,\vy) \!\ll_K\! (\vx_p,\vy_q)$. Since $\phi_t(\vx_r,\vy_s) \!\rightarrow\! (\hat \vx,\hat \vx)\in S$, and $\phi_t(\vx_p,\vy_q) \rightarrow (\bar \vx,\bar \vx)\in S$, we can once again, due to monotonicity of the system and by invoking a sandwich argument, say that $\phi_{t+\max\{t_1,t_2\}}(\vx,\vy)$ converges to an equilibrium point in $S$ as $t \!\rightarrow\! \infty$. This completes the proof.
\end{IEEEproof}

\end{document}